\documentstyle[12pt]{article}

\font\mybb=msbm10 at 11pt
\def\bb#1{\hbox{\mybb#1}}
\def\ZZ {\bb{Z}}
\def\RR {\bb{R}}

\def\V{{\cal V}}
\def\I{{\cal I}}
\def\J{{\cal J}}
\def\K{{\cal K}}
\def\C{{\cal C}}
\def\D{{\cal D}}
\def\H{{\cal H}}
\def\L{{\cal L}}

\def\G{{\cal G}}

\def\tr{{\rm tr}}
\def\exp{{\rm exp}}

\newcommand{\sectiono}[1]{\section{#1}\setcounter{equation}{0}}

\begin{document}

\thispagestyle{empty}
{\baselineskip=12pt
\hfill CALT-68-1978

\hfill hep-th/9503078

\hfill March 1995

\vspace {1.0cm}}
\parskip=9pt     
\baselineskip=17pt 
\begin{center}{\Large \bf Classical Symmetries of Some\break Two-Dimensional
Models\footnote{Work supported in part  by U.S. Department of Energy
Grant No. DE-FG03-92-ER40701.}}
\end{center}
\bigskip
\bigskip
\centerline{\large John H. Schwarz\footnote{email: jhs@theory.caltech.edu}}
\medskip
\centerline{California Institute of Technology}
\centerline{Pasadena, CA 91125}
\bigskip
\parindent=1 cm

\begin{abstract}
It is well-known that principal chiral models and symmetric space models
in two-dimensional Minkowski space
have an infinite-dimensional algebra of hidden symmetries.  Because of the
relevance of symmetric space models to duality symmetries in string theory,
the hidden symmetries of these models are explored in some
detail. The string theory application requires including coupling to
gravity, supersymmetrization, and quantum effects. However, as a first step,
this paper only considers classical bosonic theories in flat space-time.
Even though the algebra of hidden symmetries of principal
chiral models is confirmed to include a Kac--Moody algebra (or
a current algebra on a circle), it is argued that
a better interpretation is provided by a doubled current algebra on a
semi-circle (or line segment).  Neither the circle nor the semi-circle bears
any apparent relationship to the physical space. For symmetric space models
the line segment viewpoint is shown to be essential, and special boundary
conditions need to be imposed at the ends.  The algebra of hidden symmetries
also includes Virasoro-like generators. For both principal chiral
models and symmetric space models, the hidden symmetry
stress tensor is singular at the ends of the line segment.
\end{abstract}
\vfil \eject

\sectiono{Introduction}

\subsection{String theory motivation}

It is impressive how much has been learned about string theory in recent years.
Many techniques have been developed for finding large classes of classical
solutions and for computing  quantum corrections to any finite order in
perturbation theory.  It is clear that these corrections are free from
ultraviolet divergences and that the only divergences are ones that should
occur for good physical reasons.  Now there are good prospects for finding
examples that come close to making contact with the standard model and pointing
to specific phenomena, such as supersymmetry, that should be on the
experimental horizon.

It is also impressive how much is not known about string theory.  It is
not known what equation the ``classical solutions'' solve!
It is not known, even
as a matter of principle, how to compute non-perturbative quantum effects.  To
put it bluntly, we don't know the theory.  In my opinion, finding it is the
outstanding challenge in theoretical physics.  It seems unlikely that there is
a deductive path that leads to the answer.  Rather, some  inspiration seems to
be required.  Since there are many indications that
the theory will exhibit an enormous gauge
symmetry, a general strategy that could be helpful is to try to
figure out what the symmetry group should be.  Once it is identified, an
arsenal of known techniques can be brought to bear on the problem of
constructing a quantum theory with this symmetry.

This brings us to the key question:  how are we going to find the symmetry
group?  What makes this difficult is that any of the known classical solutions
involves an enormous amount of spontaneous symmetry breaking, so that the
underlying gauge symmetry is very well concealed.  This phenomenon, so familiar
from the standard model, poses a real challenge.

In recent years, there has been rapidly growing interest in various duality
symmetries of string theory.  These are of interest for a variety of reasons,
but the one that is relevant to the current discussion is that they could be
pieces of the hidden gauge symmetry.  The basic idea is that different parts of
the symmetry show up for different vacua, and that by judiciously choosing
vacua (or classical solutions) it may be possible to piece the whole story
together.  Some of these symmetries -- the $T$ dualities -- hold order by order
in string perturbation theory and are, therefore, quite well understood
\cite{bib:giveon}.  Others - such as the $S$ dualities - are inherently
nonperturbative \cite{bib:font,bib:sen,bib:schwarz}.  Therefore, the best one
can do, in the absence of the complete theory, is to find evidence that they
should be exact string theory symmetries.  Once one is satisfied of this -- as
I already am -- the issue is not to ``prove'' that they are symmetries of
string theory, which doesn't really make a lot of sense in the absence of the
theory, but rather to seek a formulation of string theory that implements them.

The string duality symmetries are infinite discrete groups.  They are best
described as subgroups of continuous groups that appear in classical
supergravity theories \cite{bib:hull}.  The restriction to the discrete
subgroups can be understood as arising from quantum corrections or from string
corrections.  The continuous symmetries are global symmetries of the classical
supergravity theory, whereas the discrete subgroup should correspond to gauge
symmetries of string theory.

The continuous global symmetries of supergravity theories are always given by
non-compact symmetric space models \cite{bib:julia,bib:salam}.  This means that
the symmetry group is a non-compact Lie group $G$, with maximal compact
subgroup $H$, and that scalar fields associated with the coset space $G/H$
(which is a non-compact symmetric space) provide a non-linear realization of
the symmetry $G$.  One class of examples is provided by toroidally
compactifying the heterotic string to $d$ dimensions and then isolating the
low-energy effective supergravity theory that describes the massless modes.
This gives a $T$ duality group $O(10-d, 26-d)$.  However, in low dimensions
there is more.  Thus, in four dimensions one also finds the $S$ duality group
$SL(2,\RR)$ \cite{bib:cremmer} and in three dimensions these merge inside a
larger group -- $O(8, 24)$ \cite{bib:marcus,bib:senb}.

In the case of the type II superstring, the effective supergravity theory is
the one with maximal supersymmetry (32 supercharges).  The group $G$ turns out
to be the maximally non-compact form of the exceptional Lie group $E_{11-d}$.
(The $E_7$ case in four dimensions was discovered first \cite{bib:cremmera}.)
This contains the $T$ duality group $O(10-d, 10-d)$ as a proper subgroup.  The
crucial fact, both in this case and the heterotic one, is that the group is
larger in lower dimensions.  In many respects, higher-dimensional field
theories represent a greater theoretical challenge.  However, when it comes to
the study of these symmetries, just the reverse is the case.  Taking the
$E_{11-d}$ literally suggests that for $d=2$ the symmetry should be $E_9$,
which is the affine extension of $E_8$.  Moreover, when all spatial dimensions
are eliminated there might even be an $E_{10}$ symmetry!  $E_{10}$ is a poorly
understood hyperbolic Lie algebra.  (See Ref. \cite{bib:gebert} for a readable
introduction.)  Its structure is rich enough that one could imagine that it is
closely related to the symmetry one is looking for as a basis for string
theory.

One feature suggested by the $E$ series is that dimensional reduction from
three dimensions to two dimensions enlarges the symmetry $G$ to $\hat G$, the
affine extension of $G$.  The purpose of this paper to explore in some detail
the extent to which this ``folk theorem'' is true in general.  (The study of
hyperbolic symmetries of one-dimensional theories is far beyond the scope of
this paper.)  Another purpose is to look for an associated Virasoro symmetry
analogous to those provided by the Sugawara construction in conformally
invariant theories.  The theories considered here are not conformally
invariant.

It is conceptually important to understand that the two-dimensional theories
studied in this paper are regarded as target-space theories, rather than
world-sheet theories.  For this reason, it makes sense to allow the spatial
dimension to be an infinite line and to discuss theories that are not
conformally invariant.  Another consequence of this viewpoint is that we do not
attach great importance to obtaining a detailed
understanding of these theories as quantum theories.
In the string theory context, high-mass string states, dropped in the massless
truncation, could make a big difference for the quantum behavior. Some
quantum issues, such as the understanding how
instanton effects pick out the discrete subgroup
should not be sensitive to details and are important for our program.

\subsection{History of hidden symmetry in two dimensions}

The study of integrable models in two dimensions possessing ``hidden'' affine
symmetries has a long and complex history.  Relativists and field theorists
were led to consider closely related models for quite different reasons.  While
the different communities had some awareness of one another, and a few
physicists belong to both of them, it is clear that more communication would
have been helpful.

The first people to discuss hidden symmetries were relativists, who were
seeking new classical solutions to the Einstein field equations in four
dimensions.  The major step that initiated this program was taken in 1971 by
Geroch \cite{bib:geroch}.  Apparently unaware of this development, field
theorists considered related systems, both classically and quantum mechanically
(but without gravity), beginning about five year later.  The main motivation
for most of them was to understand non-perturbative
properties of four-dimensional Yang--Mills theories.
They became interested in certain integrable
two-dimensional theories, a simpler prototype sharing many features with 4D
gauge theories.  The field theorists who identified
non-local charges and the associated hidden symmetries first
were  Pohlmeyer and L\"uscher \cite{bib:pohlmeyer,bib:luscher}.
Julia, whose main interest was in supergravity,
made several contributions including the clarification of
the acceptability of non-compact symmetric space models,
the discovery of a central charge in the classical hidden symmetry
algebra (when gravity is included), and the suggestion that
hyperbolic algebras should appear in one dimension \cite{bib:julia}.

In their search for classical solutions of Einstein's equations, relativists
considered systems with two Killing vectors.  If both of them are space-like
the resulting system  is described by an effective two-dimensional theory with
Lorentzian signature, whereas if one is space-like and one is time-like the
effective two-dimensional theory has Euclidean signature.  The latter case was
studied by Geroch who realized that the resulting equations were invariant
under an infinite set of transformations (the Geroch group) \cite{bib:geroch}.
He used these transformations to produce
new families of solutions from a given one.\footnote{The same technique
has been used for finding new classical
solutions to string theory.  For example, Sen has used the $S$ and $T$ duality
symmetries to construct large classes of black-hole solutions of the heterotic
string theory \cite{bib:senc}.}     He speculated that the symmetry was some
kind of loop group, but did not construct it explicitly.  Later, Kinnersley and
Chitre \cite{bib:kinnersley} constructed the infinite-dimensional Lie algebra
of infinitesimal symmetry transformations, which they identified as an affine
(or Kac--Moody) algebra.  By constructing associated linear spectral problems,
others demonstrated complete integrability \cite{bib:belinskii,bib:maisonb}.
Further clarification was provided by a reformulation as a Riemann--Hilbert
problem \cite{bib:hauser,bib:ueno}.  The inverse scattering method was
subsequently
employed by Breitenlohner and Maison \cite{bib:breitenlohner,bib:maisona}.
Other related developments can be found in Refs.
\cite{bib:ernst,bib:hausera}. Current algebra symmetries have
also been identified in two-dimensional quantum gravity
\cite{bib:polyakov,bib:knizhnik}.

Polyakov \cite{bib:polyakova} and Migdal \cite{bib:migdal} emphasized
that two-dimensional non-linear sigma models and principal chiral models share
many features with 4D gauge theories.  These include asymptotic freedom,
non-abelian symmetry groups, instantons, dynamical mass generation, and I/N
expansions.  When fermions are included axial anomalies and dynamical breaking
of $\gamma_5$ symmetry also appear.  Being much simpler theories, they were
clearly worthy objects of study.  A little later, L\"uscher and Pohlmeyer
\cite{bib:luscher} showed that non-linear sigma models have an infinite number
of conserved charges.  This was systematized by Br\'ezin, {\it et al.}
\cite{bib:brezin} and developed further in \cite{bib:zakharov}.
Similar constructions were also carried out for the
supersymmetric extensions of these models \cite{bib:curtright,bib:chaub}.
Dolan \cite{bib:dolana,bib:dolanb} showed
that the L\"uscher--Pohlmeyer charges in principal chiral models generate half
of a Kac--Moody algebra.  Her proof was simplified in Ref. \cite{bib:devchand}.
Wu \cite{bib:wub} found additional charges whose inclusion led to a complete
Kac--Moody algebra.  Another theme that arose at this time was that self-dual
Yang--Mills systems in four dimensions provide a more general setting in which
to interpret these results.  This has been pursued by Chau and collaborators
\cite{bib:chaua}, Dolan \cite{bib:dolanb}, and Atiyah \cite{bib:atiyah}
among others. Evidence has been
presented for Virasoro-like symmetries \cite{bib:cheng,bib:hou,bib:li} in
addition to the affine (or Kac--Moody) ones. More recently,
Yangian/quantum group deformations of the
Kac--Moody symmetries have been shown to arise in the quantum
sine--Gordon theory \cite{bib:bernard,bib:mackay,bib:curtrightb}.
This is somewhat puzzling to me since
we expect quantum effects to pick out discrete subgroups. In any case,
since this paper only considers classical theories, it will have nothing
to say either about such deformations or about instantons
and discrete subgroups. It should also be noted that there
is a possibility that some of the classical symmetries might not survive
quantization because of anomalies \cite{bib:goldschmidt}.

The study of hidden affine symmetries, which was initiated in the
supergravity context by Julia \cite{bib:julia},  has been advanced
considerably by Nicolai. He has made considerable progress toward
understanding the $E_9$ theory \cite{bib:nicolaia}, initiated a
study of hyperbolic symmetry algebras \cite{bib:nicolaib}, and (together with
Gebert) discussed $E_{10}$ \cite{bib:gebert}.  His review article
\cite{bib:nicolaic} is recommended.  Recently, other authors have begun to
discuss the possible relevance of this circle of ideas to string theory
\cite{bib:bakas,bib:maharana}.

Another significant development, that seems to have gone
unnoticed, was a study of the algebra of
classical hidden symmetry charges in non-linear
sigma models by Wu and by Jacques and Saint--Aubin \cite{bib:wu}.
They found
that the algebra differs from a Kac--Moody algebra in a way that they found
puzzling.  Most other authors who have addressed the
question assert incorrectly
that these models have a standard Kac--Moody symmetry algebra.

\subsection{Plan of this paper}

Section 2 discusses principal chiral models (PCM) in flat space-time.  After
setting up the formalism in any dimension,
the analysis is restricted to two-dimensional
Minkowski space using light-cone coordinates.  Then, basically following the
approach of Dolan and Wu, we derive the hidden symmetries of the model and show
that they form an affine Lie algebra.  The section concludes with a new contour
integral representation of the symmetries, which simplies and clarifies the
picture considerably.

Section 3 investigates the Virasoro symmetry of PCM's in flat
space-time.  The infinite-dimensional symmetry that appears is not quite a
Virasoro algebra, but rather a subalgebra, which we call $\V_2$.
We prove that it cannot be extended to a complete Virasoro algebra and
interpret this fact as implying that the hidden symmetry
stress tensor is singular at the ends of a line segment.

Section 4 describes a non-linear non-local change of variables that is
interpreted as a non-abelian duality transformation. (This section is
not required for understanding Section 5.) The transformed theory
has a Wess-Zumino term.  We argue that any coefficient $u$ of the WZ term
other than $u = \pm 1$, which would give Wess-Zumino-Novikov-Witten theory
\cite{bib:wess,bib:novikov,bib:wittenb},
is equivalent to $u = 0$. A brief comment about the quantum theory is made at
this point.  The hidden
symmetry of the $u$-transformed theory is derived directly.

Section 5 analyzes symmetric space models (SSM) $G/H$ in flat two-dimensional
Minkowski space-time.  After describing the general formalism, we explore the
hidden symmetries and work out their algebra.  As we have already indicated, we
do not find a standard Kac--Moody algebra $\hat G$, but rather a new structure,
which we name $\hat G_H$.  It is interpreted as a group on a line segment,
rather than a circle, subject to certain boundary conditions at the ends of the
line.  The distinction is reminiscent of the difference between closed strings
and open strings, though that is only an analogy.  The Virasoro symmetry is
shown to be defined on the same line segment, also with specific boundary
conditions at the ends. A check of the results is provided by
examining the formulation of PCM's as SSM's.

\sectiono{Principal Chiral Models in Flat Space-time}
\subsection{Basic formulas}

We begin by reviewing the definition of a principal chiral model (PCM) for a
Minkowski space-time of arbitrary dimension.  The basic variables are
matrices $g(x)$ which map the space-time into a particular representation $R$
of a Lie  group $G$.  All considerations in this paper are purely classical.
Indeed the goal is to determine the invariance group of the
classical equations of motion.
Let ${\cal G}$ denote the Lie algebra corresponding to $G$ and let
$T_i$ be a basis
for ${\cal G}$ in the representation $R$
\begin{equation}
[T_i,T_j] = f_{ij}{}^k T_k.
\end{equation}

The classical PCM is then defined by the lagrangian
\begin{equation}
{\cal L} = \eta^{\mu\nu} tr (A_\mu A_\nu),
\end{equation}
where $\eta^{\mu\nu}$ denotes the Minkowski metric and
\begin{equation}
A_\mu = g^{-1} \partial_\mu g = \sum A_\mu^i T_i.
\end{equation}
Since our considerations are entirely classical, a
normalization factor in ${\cal L}$ has been omitted.
${\cal L}$ can also be written in the form
\begin{equation}
{\cal L} = \eta^{\mu\nu} tr (\tilde{A}_\mu \tilde{A}_\nu),
\end{equation}
where
\begin{equation}
\tilde{A}_\mu = g \partial_\mu g^{-1} = - \partial_\mu g \cdot g^{-1} = - g
A_\mu g^{-1} = \sum \tilde{A}_\mu^i T_i.
\end{equation}
Let $\delta g$ represent an arbitrary infinitesimal variation of $g$.  Then,
since $g^{-1} \delta g$ belongs to the Lie algebra ${\cal G}$, we can write
\begin{equation}
\delta g = g\eta
\end{equation}
where $\eta = \sum \eta^i T_i$. Under this variation
\begin{equation}
\delta A_\mu = D_\mu \eta = \partial_\mu \eta + [A_\mu, \eta],
\end{equation}
and thus
\begin{equation}
\delta {\cal L} = 2\, {\rm tr} (A^\mu D_\mu \eta) = 2\, {\rm tr}
(A^\mu \partial_\mu \eta).
\end{equation}
{}From this we conclude that the classical equation of motion is
\begin{equation}
\partial_\mu A^\mu = 0\quad {\rm or} \quad \partial_\mu \tilde{A}^\mu = 0.
\end{equation}
Since $A_\mu$ is pure gauge,
\begin{equation}
F_{\mu\nu} = \partial_\mu A_\nu - \partial_\nu A_\mu + [A_\mu, A_\nu] = 0
\end{equation}
is the Bianchi identity, though it is usually called the Cartan--Maurer
equation in this context.
It is evident that the transformation $\delta g$, with $\eta$ constant, leaves
${\cal L}$ and the equation of motion invariant.  A similar symmetry described
by left multiplication is obtained from the corresponding reasoning based on
$\delta g = - \tilde{\eta} g$ and the $\tilde{A}$ formulation of ${\cal L}$.
Thus, the PCM in any dimension has global $G \times G$ symmetry.

Half of the $G\times G$ symmetry can be regarded as a
gauge symmetry and used to effect a gauge choice, such as $g (x_0)=1$,
where $x_0^\mu$ is a fixed point in space-time. This can be achieved
by making a change of variables $ g(x) \to g(x) g(x_0)^{-1}$.
This may not be desirable, however, since it effectively removes the
zero mode degree of freedom. In the subsequent discussion such a
``choice of gauge'' will  be used on occasion as a way of simplifying
certain equations. This is not essential, however, and none of the
results will depend on doing this.
In the case of a PCM, the left or right group action $(\delta g = g\eta$ or
$\delta  g = - \eta g)$ is the non-abelian counterpart of the translation
symmetry of a free massless scalar.  Either one of these symmetries can be used
to choose the gauge $g (x_0) = 1$.  Once this is done, the remaining non-gauge
symmetry, which is also $G$, is described by $\delta g = [g,\eta]$,
since this is the unique combination that preserves the gauge choice.
If we set $g = 1$ for
a particular space-time point $x_0^\mu = (x_0^0, x_0^i)$, the
Hamiltonian evolution of the system, $\dot g = i [ H,g]$, preserves the
gauge.  In other words, $g (x^0, x_0^i) = 1$, for all $x^0$.  This is analogous
to the preservation of the Gauss law constraint in electrodynamics.  (Quantum
mechanically, one should only require that the identity holds for matrix
elements between physical states.) In the case of two dimensions,
many authors who have written on
this subject require that $g=1$ at $x^1 = -\infty$, which is a particular
example of this type of gauge choice.

The main purpose of this manuscript is to describe the extension of the $G
\times G$ symmetry that occurs in two dimensions.  For this purpose it is very
convenient to introduce light-cone coordinates
\begin{equation}
x^\pm = x^0 \pm x^1,\quad \partial_\pm = {1\over 2} (\partial_0 \pm
\partial_1),
\end{equation}
so that the equation of motion and Bianchi identity take the forms
\begin{equation}
\partial_{\mu} A^{\mu} = \partial_+ A_- + \partial_- A_+ = 0
\end{equation}
\begin{equation}
F_{+-} = \partial_+ A_- - \partial_- A_+ + [A_+, A_-] = 0.
\end{equation}
These can be solved to give
\begin{equation}
\partial_- A_+ = - \partial_+ A_- = {1 \over 2} [A_+ , A_- ] .
\label{eq:laxsolve}
\end{equation}
This theory should be contrasted with $WZNW$ theory, which has a Wess--Zumino
term added in such a way that the equation of motion becomes $\partial_+ A_- =
0$, which is equivalent to $\partial_- \tilde{A}_+ = 0$.  In this case it is
well-known and easy to see that the symmetry becomes $\hat G \times \hat G$,
where $\hat G$ denotes an affine -- or current algebra --
extension of $G$. Classically the algebra $\hat G$ has no central term,
but quantum mechanically
it has one controlled by the normalization of ${\cal L}$.  In the case of a PCM
in 2D, without Wess-Zumino term, the occurrence of an affine extension of the
symmetry is much less transparent, and it is our purpose to explain in detail
how it arises.

\subsection{Half of the symmetry}

A standard technique for discovering the ``hidden symmetries'' of integrable
models such as the 2D PCM begins by considering a pair of
equations, known as a Lax pair \cite{bib:lax}.
This is a pair of differential equations for an auxiliary
group-valued quantity,
which we call $X$. It depends on the field $g(x)$, the
space-time coordinate $x$, a basepoint $x_0$, and an
additional ``spectral parameter'' $t$. The quantity $X$ will play a
central role in all subsequent descriptions of hidden symmetries.
The compatibility of the Lax pair is arranged to
encode the conditions one wants to implement, which in our case are
$\partial\cdot A = F_{+-}=0.$ The Lax pair that is appropriate to our problem
is
\begin{equation}
(\partial_+ + \alpha_+ A_+) X = 0 \quad {\rm and} \quad
(\partial_- + \alpha_- A_-) X = 0,  \label{eq:laxpair}
\end{equation}
where $\alpha_\pm$ are constants.  Requiring these equations to be compatible
(using $\partial\cdot A = F_{+-}=0$) gives the condition
$\alpha_+ + \alpha_- = 2 \alpha_+ \alpha_-$.  Therefore, the general solution
is given in terms of the spectral parameter by
\begin{equation}
\alpha_+ = {t\over t - 1} , \quad \alpha_- = {t\over t + 1}.
\end{equation}
Rescaling the differential operators gives an equivalent Lax pair
\begin{equation}
(\partial_\pm \mp t D_\pm) X = 0 ,
\end{equation}
whose compatibility equation is
\begin{equation}
[\partial_- + tD_-, \partial_+ - tD_+] = - t \partial \cdot A + t^2 F_{+-} = 0.
\end{equation}

A formal solution to the Lax pair is given by
\begin{equation}
X (x^\mu, t) = P \exp \Big\{ - \int_{x_{0}}^x (\alpha_+ A_+ dy^+
+ \alpha_- A_- dy^-)\Big\},
\end{equation}
where the integration is along an arbitrary contour from $x_0^\mu$ to $x^\mu$
and $P$ denotes path ordering with the $x$ end
of the contour on the left and the $x_0$ end on
the right.  Note that $X$ takes values in the Lie group, and that the
expression
is independent of the choice of
contour provided that the equation of motion is
satisfied and the space-time is simply connected.  (An interesting
problem, which will not be considered here, is to extend the analysis to a
circular spatial coordinate.)  Now $X$, as given, is only well-defined on
shell.  However, by committing to a specific prescription for the integration
contour it can be extended off shell.  This has been done by most previous
authors, but will not be done here, since the formula is only
required on shell, and we prefer to avoid unnecessary arbitrary choices.
Once a particular off-shell prescription is chosen, the variation of the
Lagrangian can be examined, though we will not do so here.
The invariance of the equations of motion
guarantees that it will be a total derivative, whose
form depends on the prescription.

As an aside, let us remark that eqs.(\ref{eq:laxpair}) can be
rewritten in the form
\begin{equation}
{\hat A}_{\pm} (x,t) = X \partial _{\pm} X^{-1} = \alpha_{\pm}(t) A_{\pm}(x).
\label{eq:ahat}
\end{equation}
Then eqs.(\ref{eq:laxsolve}) and (\ref{eq:ahat}) imply that
\begin{equation}
{\hat F}_{+ -} = \partial_+ {\hat A}_- - \partial_- {\hat A}_+
+[{\hat A}_+ , {\hat A}_-] = \big(\alpha_+ \alpha_-
- {1 \over 2} (\alpha_+ + \alpha_-)\big) [A_+, A_-] = 0.
\end{equation}

The next step is to define an infinitesimal matrix-valued function
\begin{equation}
\eta (\epsilon, t) = X(t) \epsilon X(t)^{-1},
\end{equation}
where $\epsilon = \sum \epsilon^i T_i$ and $\epsilon^i$ are infinitesimal
constants.  The claim, then, is that the infinitesimal transformation
$\delta(\epsilon,t) g = g \eta $ preserves the
equation of motion.  This is the ``hidden symmetry'' that we intend to explain.
Before presenting the proof, let us first
explain the notation carefully, since it could be confusing otherwise.
First of all, the symbols $\delta$ and $\Delta$ are always used to designate
operators that implement infinitesimal variations in contrast to the symbols
$\epsilon$ and $\eta$, which denote infinitesimal matrices
and matrix-valued functions, respectively.
Since the formulas contain an arbitrary ``spectral parameter''
$t$ and are analytic in a neighborhood of $t=0$,
we may expand the variation in a power series
\begin{equation}
\delta (\epsilon,t) = \sum_{n = 0}^\infty t^n \delta_n (\epsilon)
\end{equation}
to define an infinite number of distinct global symmetry transformations:
\begin{eqnarray}
g^{-1} \delta_0 (\epsilon) g &=& \epsilon\\
g^{-1} \delta_1 (\epsilon) g &=& [\int_{x_{0}}^x (A_+ dy^+ - A_- dy^-),
\epsilon] \label{eq:iappears}
\end{eqnarray}
and so forth.  The $\delta_n$ transformations for $n > 0$ are generated by the
celebrated L\"uscher--Pohlmeyer non-local charges.  The most general
variation associates a distinct infinitesimal parameter
$\epsilon_n$ to each of the $\delta_n$'s.
The formulas involving $\epsilon_n$'s can be obtained
from $\delta(\epsilon,t)$ by replacing $\epsilon t^n$ by $\epsilon_n$.
This replacement can be implemented by contour integration in an obvious way.

The proof of the hidden symmetry is now easy.
First note that the Lax pair implies that
\begin{equation}
D_\pm \eta = \partial_\pm \eta + [A_\pm, \eta] = \pm {1\over t} \partial_\pm
\eta.  \label{eq:deta}
\end{equation}
Therefore,
\begin{equation}
\delta (\partial \cdot A) = \partial_+ (D_- \eta) + \partial_- (D_+ \eta) = 0,
\end{equation}
as required.

The conserved currents associated with the hidden symmetry are given by
$D_\mu \eta$.  Therefore, the corresponding conserved charges are
\begin{equation}
Q (\epsilon, t) = \int_{-\infty}^\infty D_0 \eta \, dx^1 = {1\over t}
\int_{-\infty} ^\infty \partial_1 \eta\,  dx^1
 = {1\over t} \eta \Big|_{-\infty}^\infty ,
\end{equation}
where we have used eq. (\ref{eq:deta}).  In particular,
for the choice $x_0^1 = - \infty$
(which many authors prefer)
\begin{equation}
Q^i (t) = {1\over t} X_\infty (t) T^i X_\infty (t)^{-1},
\end{equation}
where
\begin{equation}
X_\infty (t) = P \exp \Big\{- \int_{-\infty}^\infty
(\alpha_+ A_+ - \alpha_- A_-)dx^1\Big\}
\end{equation}
is time independent.  ($t$ is the spectral parameter.)  If the
spatial dimension were a circle, rather than an infinite line, $X_\infty (t)$
would be replaced by the corresponding Wilson loop.  This would still require
the choice of a base point $x_0$.

The conserved charges generate the symmetry transformations, so
\begin{equation}
[Q(\epsilon, t), g] = \delta (\epsilon, t) g.
\end{equation}
In the classical theory, which is all that is considered in this paper, the
bracket is a Poisson bracket.  Nonetheless, we will usually refer to it as a
commutator, since the Poisson bracket of two charges generates the commutator
of the corresponding symmetry transformations.

\subsection{Half of the algebra}

Here we wish to examine the effect of commuting two of the symmetry
transformations.  In other words we wish to compute
$[\delta (\epsilon_1, t_1), \delta (\epsilon_2, t_2)] g(x)$.
To save writing we let $\delta_i = \delta (\epsilon_i, t_i), X_i = X(t_i)$, and
$\eta_i = \eta (\epsilon_i, t_i) = X_i \epsilon_i X_i^{-1}$.  In this notation,
we want to compute $[\delta_1, \delta_2]g$.  Commuting transformations in this
way is equivalent to computing the Poisson bracket of the corresponding
conserved charges.\footnote{It was pointed out to me by C. Hull that this
procedure can fail to detect a central term in the Poisson bracket
algebra. That does not appear to happen for the cases considered
in this paper.}
In doing this we will use
the equations of motion.  This is not so bad, since generically, for any
off-shell extension of the formulas, the commutators will contain irrelevant
extra terms that vanish when the equations are imposed.  Since these terms are
not of much interest, we might as  well impose the equation of motion in the
first place.  Our purpose, after all, is to determine the group of
transformations that preserves the equation of motion.

The key formula that makes it possible to obtain an elegant result for the
algebra, in the  notation described above, is
\begin{equation}
\delta_1 X_2 = {t_2\over t_1 - t_2} (\eta_1 X_2 - X_2 \epsilon_1).
\label{eq:deltax}
\end{equation}
Letting $Q$ denote either side of this equation, the method of proof is to
first note that both sides satisfy the boundary condition
$Q(x_0) = 0$, which is obvious since $\eta_1 (x_0) =
\epsilon_1$ and $X_2 (x_0) = 1$.  The less trivial step is to verify that both
sides satisfy the pair of differential equations
\begin{equation}
(\partial_{\pm} + \alpha_{2\pm} A_{\pm}) Q
+ \alpha_{2\pm} (D_{\pm}\eta_1) X_2 =0.
\end{equation}
In the case of $Q = \delta_1 X_2$, these equations arise from varying the Lax
pair.  For the
right-hand expression for $Q$, the equation is verified by direct substitution
and use of the Lax pair.  These equations plus the boundary condition at $x_0$
are sufficient to deduce the desired identity.

Now we are ready to compute
\begin{equation}
[\delta_1, \delta_2] g = g (\delta_1 \eta_2 - \delta_2 \eta_1 + [\eta_1,
\eta_2]).
\end{equation}
The equation just obtained for $\delta_1 X_2$ implies that
\begin{equation}
\delta_1 \eta_2 = {t_2\over t_1 - t_2} ([\eta_1, \eta_2] - X_2 \epsilon_{12}
X_2^{-1}),
\end{equation}
where $\epsilon_{12} = [\epsilon_1, \epsilon_2] = f_{ij}{}^k
\epsilon_1^i \epsilon_2^j T_k$.  Thus
\begin{equation}
\delta_1 \eta_2 - \delta_2 \eta_1 + [\eta_1, \eta_2] = {t_1 X_1 \epsilon_{12}
X_1^{-1} - t_2 X_2 \epsilon_{12} X_2^{-1}\over t_1 - t_2}.
\end{equation}
This then implies that
\begin{equation}
[\delta (\epsilon_1, t_1), \delta (\epsilon_2, t_2)] = {t_1 \delta
(\epsilon_{12}, t_1) - t_2 \delta (\epsilon_{12}, t_2)\over t_1 - t_2},
\label{eq:pcmalgebra}
\end{equation}
which is the main result.  Expanding in power series, this gives
\begin{equation}
[\delta_m (\epsilon_1), \delta_n (\epsilon_2)]
= \delta_{m + n} (\epsilon_{12}), \quad m,n \geq 0
\end{equation}
which is half of an affine current algebra.

\subsection{The rest of the symmetry}

Since $g \rightarrow g^{-1}$ is an automorphism of a PCM, we can define a
second set of symmetry transformations by
$\tilde{\delta} (\epsilon, t) g^{-1} = g^{-1} \tilde{\eta} (\epsilon, t)$
or, equivalently,
\begin{equation}
\tilde{\delta} (\epsilon, t) g = - \tilde{\eta} (\epsilon, t) g,
\end{equation}
where
\begin{equation}
\tilde{\eta} (\epsilon, t) = \tilde{X} (t) \epsilon \tilde{X} (t)^{-1}
\end{equation}
and
\begin{equation}
\tilde{X} (t) = P \exp \Big\{ - \int_{x_{0}}^x (\alpha_+ \tilde{A}_+ dy^+ +
\alpha_- \tilde{A}_- dy^-)\Big\}.
\end{equation}
Equation (\ref{eq:pcmalgebra}) and the automorphism then imply that
\begin{equation}
[\tilde{\delta} (\epsilon_1, t_1), \tilde{\delta} (\epsilon_2, t_2)] =
{t_1 \tilde{\delta} (\epsilon_{12}, t_1) - t_2 \tilde{\delta}
(\epsilon_{12}, t_2)\over t_1 - t_2},
\end{equation}
which also gives half of an affine current algebra.
Unfortunately $[\delta (\epsilon_1, t_1),
\tilde{\delta} (\epsilon_2, t_2)]$ is a little messy,
so it is not clear at this
point what the complete algebra becomes.  Circumventing this difficulty
requires another nice formula.

Let us define
\begin{equation}
Y(t) = X (1/t) = P\exp \left\{\int_{x_{0}}^x \left({1\over t-1} A_+ dy^+ -
{1\over t+1} A_- dy^-\right)\right\}
\end{equation}
and consider another transformation formula:
\begin{equation}
\bar \delta (\epsilon,t) g = - g Y(t) \epsilon Y(t)^{-1}.
\end{equation}
Aside from a minus sign, this is the ``analytic continuation'' of the $\delta$
transformation.  However, expanding this formula in powers of $t$
\begin{equation}
\bar  \delta (\epsilon,t) = \sum_{n=0}^\infty t^n \bar \delta_n (\epsilon)
\label{eq:deltabar}
\end{equation}
corresponds to an expansion about $t = \infty$ in the previous formula, which
may seem dangerous since there are singularities at $t = \pm 1$.  However,
$Y(t)$ can be recast in another form which makes everything clear:
\begin{equation}
Y(t) = g^{-1} (x) \tilde{X} (t) g_0,
\end{equation}
where $g_0 = g (x_0)$.  This formula is proved by the same method that we used
to derive $\delta_1 X_2$.  It is easy to show that both sides are 1 when
evaluated at $x = x_0$, and they both satisfy the pair of equations
$(\partial_{\pm} + \alpha_{\pm}(t^{-1}) A_{\pm})Y =0$.
This is sufficient to prove the formula.  Substituting this result gives
\begin{equation}
\bar{\delta} (\epsilon, t) g = \tilde{\delta} (g_0 \epsilon g_0^{-1}, t) g.
\end{equation}
Thus $\bar{\delta}$ is the same as a $\tilde{\delta}$ transformation with
the parameter conjugated by the constant group element $g_0$.  Symmetry under
$\bar \delta$ and $\tilde{\delta}$ are, therefore, completely equivalent.
However, their algebras are not the same.  The reason for this is that, even
though $g_0$ is a constant, it does transform:
\begin{equation}
\delta (\epsilon, t) g_0 = g_0 \epsilon \quad {\rm and} \quad
\bar\delta (\epsilon, t) g_0 = - g_0 \epsilon.
\end{equation}

Whereas $[\delta, \tilde{\delta}]$ is difficult to compute, $[\delta,
\bar\delta]$ is not hard.  Indeed, the correct result is obtained by the
substitution
\begin{equation}
\bar\delta (\epsilon,t) = - \delta (\epsilon, {1\over t})
\end{equation}
in the $[\delta, \delta]$ algebra.  Of course, the reader should be wary of the
validity of this formal trick.  The real proof is to compute $[\delta,
\bar\delta]$ and $[\bar\delta, \bar\delta]$ by the same methods that we used
for $[\delta, \delta]$.  In any case, one learns that
\begin{equation}
[\delta (\epsilon_1, t_1), \bar\delta (\epsilon_2, t_2)] = {t_1 t_2 \delta
(\epsilon_{12}, t_1) + \bar\delta (\epsilon_{12}, t_2)\over 1 - t_1 t_2},
\end{equation}
which implies that
\begin{equation}
 [\delta_m (\epsilon_1), \bar\delta_n (\epsilon_2)] = \left\{\begin{array}{ll}
\delta_{m-n} (\epsilon_{12}) & m > n > 0 \\
\bar\delta_{n - m} (\epsilon_{12}) & n > m \geq 0 \\
\delta_0 (\epsilon_{12}) + \bar\delta_0 (\epsilon_{12}) & m = n > 0 \\
\bar\delta_0 (\epsilon_{12}) \delta_{m,0} & n = 0 \quad . \end{array} \right.
\end{equation}
(Note that the symbol $\delta_{m,0}$ is a Kronecker $\delta$.)
Similarly,
\[ [\bar\delta (\epsilon_1, t_1), \bar\delta (\epsilon_2, t_2)] = {t_2
\bar\delta (\epsilon_{12}, t_1) - t_1 \bar\delta (\epsilon_{12}, t_2)\over t_1
- t_2}\]
\begin{equation}
= {t_1 \bar\delta (\epsilon_{12}, t_1) - t_2 \bar\delta (\epsilon_{12},
t_2)\over t_1 - t_2} - \bar\delta (\epsilon_{12}, t_1) - \bar \delta
(\epsilon_{12}, t_2),
\end{equation}
which implies that
\begin{equation}
[\bar\delta_m (\epsilon_1), \bar\delta_n (\epsilon_2)] = \bar\delta_{m+n}
(\epsilon_{12}) - \bar\delta_m (\epsilon_{12}) \delta_{n,0} - \bar\delta_n
(\epsilon_{12}) \delta_{m,0} .
\end{equation}

All of the preceding results can be elegantly reassembled by defining
\begin{equation}
\Delta_m (\epsilon) = \delta_m (\epsilon) \quad ,
\quad \Delta_{-m} (\epsilon) = \bar\delta_m (\epsilon) \quad m>0
\label{eq:deltaall}
\end{equation}
\begin{equation}
\Delta_0 (\epsilon) = \delta_0 (\epsilon) + \bar\delta_0 (\epsilon)
\quad , \quad
\bar\Delta (\epsilon) = - \bar\delta_0 (\epsilon).
\end{equation}
Then the complete algebra becomes
\begin{equation}
[\Delta_m (\epsilon_1), \Delta_n (\epsilon_2)] = \Delta_{m+n} (\epsilon_{12})
\qquad m,n \in {\ZZ}
\end{equation}
\begin{equation}
[\Delta_m (\epsilon_1), \bar\Delta (\epsilon_2)] = 0
\end{equation}
\begin{equation}
[\bar\Delta (\epsilon_1), \bar\Delta (\epsilon_2)] = \bar\Delta
(\epsilon_{12}).
\end{equation}
This describes a symmetry group $\hat G \times G$, where now $\hat G$ is a
complete affine current algebra (without center),
also called a loop algebra or Kac--Moody algebra.

Finally, let us see how the original global $G \times G$ symmetry is
realized here:
\begin{equation}
\Delta_0 (\epsilon) g = (\delta_0 (\epsilon) + \bar\delta_0 (\epsilon)) g
= g \epsilon - g_0 \epsilon g_0^{-1} g
\end{equation}
\begin{equation}
\bar\Delta (\epsilon) g = - \bar\delta_0 (\epsilon) g= g_0 \epsilon g_0^{-1}
g.
\end{equation}
In particular,
\begin{equation}
\Delta_0 (\epsilon) g_0 = 0,\quad \bar\Delta (\epsilon) g_0 = g_0 \epsilon.
\end{equation}
Using these formulas,
it is easy to verify that $[\Delta_0 (\epsilon), \bar\Delta
(\epsilon)]g = 0$. The $g_0$ factors are a bit ugly,
but they can be eliminated by
choosing the $g_0 = 1$ gauge.  The way to do this is to make a change of
variables $g' = g_0^{-1} g$.  It then follows that
\begin{equation}
\Delta_0 (\epsilon) g' = [g', \epsilon]
\quad {\rm and} \quad \bar\Delta (\epsilon) g' = 0.
\end{equation}
The interpretation of this result is that the factor of $G$ in $\hat G \times
G$ is a gauge symmetry, which disappears  when we fix the gauge $g_0 = 1$.
Then only the affine algebra $\hat G$ survives as a physical
(non-gauge) symmetry.  Note
that the $\Delta_0$ transformation in the $g_0=1$ gauge preserves the gauge.
Henceforth, when we want to fix a gauge, we will simply set $g_0 = 1$.
Alternatively, if one does not want to fix a gauge, one can get from
$\delta$ to $\bar\delta$ directly by using the the $g_0$ preserving
automorphism $g \to g_0 g^{-1} g_0$.

\subsection{A check of the algebra}

In the discussion above the commutator $[\delta (\epsilon_1, t_1), \bar\delta
(\epsilon_2, t_2)]$ was obtained by a procedure of questionable rigor.  One
could imagine, for example, that due to some subtlety it misses a central term
in the algebra.  If that were to happen, one would have
\begin{equation}
[\Delta_m (\epsilon_1), \Delta_n (\epsilon_2)] = \Delta_{m+n} (\epsilon_{12}) +
m\delta_{m+n,0} tr (\epsilon_1 \epsilon_2) Z,
\end{equation}
for some suitable number or operator $Z$, which is central, {\it i.e.},
$[Z, \Delta_m (\epsilon)] = 0$.

To verify that the formulas given above are correct without a
central term, let us choose the $g_0=1$ gauge and compute
$[\Delta_1 (\epsilon_1), \Delta_{-1} (\epsilon_2)]g$.  The question is whether
or not the result is $\Delta_0 (\epsilon_{12})g = [g, \epsilon_{12}]$.  To do
this calculation, we need that
\begin{equation}
\Delta_1 (\epsilon_1) g = \delta_1 (\epsilon_1) g = g [ \int_{x_{0}}^x (A_+
dy^+ - A_- dy^-), \epsilon_1]
\end{equation}
\begin{equation}
\Delta_{-1} (\epsilon_2) g = \bar\delta_1 (\epsilon_2)g = - [ \int_{x_{0}}^x
(\tilde{A}_+ dy^+ - \tilde{A}_- dy^-), \epsilon_2]g.
\end{equation}
{}From these it follows that
\begin{equation}
\Delta_1 (\epsilon_1) \tilde{A}_\pm  = \mp \partial_\pm (g \epsilon_1 g^{-1})
\end{equation}
\begin{equation}
\Delta_{-1} (\epsilon_2) A_\pm = \mp \partial_\pm (g^{-1} \epsilon_2 g)
\end{equation}
Thus, a straightforward calculation gives
\begin{equation}
[\Delta_1(\epsilon_1), \Delta_{-1} (\epsilon_2) ]g
= [ \int_{x_{0}}^x \partial_\mu (g \epsilon_1 g^{-1})dy^\mu, \epsilon_2]g
+ g [ \int_{x_{0}}^x \partial_\mu (g^{-1} \epsilon_2 g) dy^\mu,
\epsilon_1] = [g, \epsilon_{12}]
\end{equation}
as expected.  Thus the algebra does not have a central term of the
type that can be detected in this way. (See footnote 4.)

\subsection{The messy commutator}

In the preceding discussion we remarked that the commutator
\begin{equation}
[\delta_1, \tilde{\delta}_2] g = [\delta (\epsilon_1, t_1), \tilde{\delta}
(\epsilon_2, t_2)]g
\end{equation}
is somewhat messy.  By introducing $\bar\delta$
instead of $\tilde{\delta}$ we
were able to avoid confronting this problem and
to derive a complete affine
symmetry algebra.  In Section 5.4, for reasons that
would be difficult to explain
at this point, we will want to know $[\delta_1,
\tilde{\delta}_2]$.  Therefore,
we compute it now.  From the preceding discussion,
\begin{equation}
\tilde{\delta} (\epsilon_2, t_2) = \bar\delta (g_0^{-1}
\epsilon_2 g_0, t_2) =
- \delta (g_0^{-1} \epsilon_2 g_0, {t_2^{-1}}).
\end{equation}
Therefore,
\begin{equation}
[\delta_1, \tilde{\delta}_2] g = - [\delta (\epsilon_1, t_1), \delta (g_0^{-1}
\epsilon_2 g_0, {t_2^{-1}})]g.
\end{equation}
This commutator has two contributions.  One comes
from the general formula (\ref{eq:pcmalgebra})
for the commutator of two $\delta$'s.  The other comes
from the variation of the $g_0$'s in the argument of the second transformation.
 Using $\delta_1 g_0 = g_0 \epsilon_1$, we have
\begin{equation}
\delta_1 (g_0^{-1} \epsilon_2 g_0) = - \epsilon'_{12},
\end{equation}
where
\begin{equation}
\epsilon'_{12} = \epsilon_1 g_0^{-1} \epsilon_2 g_0 - g_0^{-1} \epsilon_2 g_0
\epsilon_1.
\end{equation}
Thus, we obtain
\[ [\delta_1, \tilde{\delta}_2] g = \left(\delta(\epsilon'_{12},
{t_2^{-1}}) - {t_1
\delta(\epsilon'_{12}, t_1) - { t_2}^{-1} \delta(\epsilon'_{12},
{t_2^{-1}})\over
t_1 - {t_2^{-1}}}\right) g \]
\begin{equation}
= {t_1 t_2\over 1-t_1 t_2} (\delta(\epsilon'_{12}, t_1) -
\delta(\epsilon'_{12}, {t_2^{-1}}))g
= {t_1 t_2\over 1 - t_1 t_2} (\delta(\epsilon'_{12}, t_1) + \tilde{\delta}
(g_0 \epsilon'_{12} g_0^{-1}, t_2))g,
\end{equation}
which is the desired formula.

The derivation of this result involves one step that could be questioned.
Namely, one could imagine that the implicit analytic continuations that are
required to interpret variations involving reciprocals of spectral parameters
could introduce subtle errors.  Since it will be very important to us  in
Section 5.4 to have confidence in the formula, we now give a second derivation
that does not require analytic continuation.  Direct computation
gives
\begin{equation}
[\delta_1, \tilde{\delta}_2] g = - \delta_1 (\tilde{X}_2 \epsilon_2
\tilde{X}_2^{-1}) g - g \tilde{\delta}_2 (X_1 \epsilon_1 X_1^{-1}).
\end{equation}
This can be evaluated once we know formulas for $\delta_1 \tilde{X}_2$ and
$\tilde{\delta}_2 X_1$.  This is the same sort of problem we encountered
previously (in the case of $\delta_1 X_2$) and the method of solution is the
same.  Variation of the Lax pair gives
\begin{equation}
(\partial_\pm + \alpha_{2\pm} \tilde{A}_\pm) \delta_1 \tilde{X}_2 +
\alpha_{2\pm} (\delta_1 \tilde{A}_\pm) \tilde{X}_2 = 0.
\end{equation}
Using the identity
\begin{equation}
\partial_\pm (g \eta_1 g^{-1}) = (\alpha_{1\pm} - 1) [\tilde{A}_\pm, g\eta_1
g^{-1}],
\end{equation}
one can show that
\begin{equation}
\delta_1 \tilde{A}_\pm = - D_\pm (g \eta_1 g^{-1}) = - \alpha_{1\pm}
[\tilde{A}_\pm, g \eta_1 g^{-1}].
\end{equation}
Therefore, $\delta_1 \tilde{X}_2$ is determined by the differential equations
\begin{equation}
(\partial_\pm + \alpha_{2\pm} \tilde{A}_\pm) \delta_1 \tilde{X}_2 =
\alpha_{1\pm} \alpha_{2\pm} [\tilde{A}_\pm, g\eta_1 g^{-1}]\tilde{X}_2,
\end{equation}
subject to the boundary condition $\delta_1 \tilde{X}_2 |_{x_{0}} = 0$.  The
unique solution to this is easily found to be
\begin{equation}
\delta_1 \tilde{X}_2 = {t_1 t_2\over 1 - t_1 t_2} (\tilde{X}_2 g_0 \epsilon_1
g_0^{-1} - g \eta_1 g^{-1} \tilde{X}_2).
\end{equation}
Similarly, one finds that
\begin{equation}
\tilde{\delta}_2 X_1 = {t_1 t_2\over 1 -t_1 t_2} (X_1 g_0^{-1} \epsilon_2 g_0 -
g^{-1} \tilde{\eta}_2 g X_1).
\end{equation}
Substituting these expressions in the formula for $[\delta_1, \tilde{\delta}_2]
g$ above gives the same result as before, namely
\begin{equation}
[\delta_1, \tilde{\delta}_2] g = {t_1 t_2\over 1 - t_1 t_2} \Big(\delta
(\epsilon'_{12}, t_1) + \tilde{\delta}
(g_0 \epsilon'_{12} g_0^{-1}, t_2 )\Big) g.
\label{eq:tildecom}
\end{equation}
Therefore, we can invoke this result with confidence in Section 5.4.

\subsection{Contour integral representation}

When $\delta (\epsilon, t) g$ is analytically continued throughout
the complex $t$-plane
the only singularities encountered are at $t = \pm 1$.
Since these singularities involve poles of unbounded order they are
essential singularities.  However, they do not introduce branch
cuts.  Let ${\cal C}_\pm$ denote small {\it clockwise}
contours in the $t$-plane about
$\pm 1$, such that the point $t=0$ is outside both of them,
and let ${\cal C} = {\cal C}_+ + {\cal C}_-$.  Then consider
\begin{equation}
\Delta_n (\epsilon) g = \int_{\cal C} {dt\over 2\pi i} t^{-n -1}
\delta (\epsilon, t) g  \quad  n \in \ZZ. \label{eq:deltan}
\end{equation}
These integrals are
well-defined finite expressions, and
they obviously describe symmetries of the theory, since $\delta
(\epsilon, t)$ does.

It is instructive to examine what happens to the
right-hand side of eq. (\ref{eq:deltan})
under a change of variable $t \rightarrow 1/t$.  Since $\C_+$ and $\C_-$ are
essentially unchanged, this gives
\begin{equation}
\Delta_n (\epsilon) g = - \int_{\C}
{dt\over 2\pi i} t^{n-1} \delta (\epsilon, 1/t) g.
\end{equation}
The Jacobian gives the minus sign, because $\C$ does not enclose the
origin.  Recall that in Section 2.4 we introduced an operation $\bar\delta
(\epsilon, t)g$ with the property that
\begin{equation}
\delta (\epsilon, 1/t) g = - \bar\delta (\epsilon, t) g.
\end{equation}
Therefore
\begin{equation}
\Delta_{-n} (\epsilon) g = \int_\C {dt\over 2\pi i} t^{-n - 1} \bar\delta
(\epsilon, t) g,
\end{equation}
which is consistent with eqs. (\ref{eq:deltabar}) and (\ref{eq:deltaall}).
Note that if we had included an
extra function $f(t)$ in the integral defining $\Delta_n (\epsilon) g$, it
would violate the required symmetry unless $f(t) = f (1/t)$.  In other words,
the symmetry alone determines the integral up to a function of $t + t^{-1}$.

One consequence of this structure for the contour integrals is that when we
evaluate commutators of the form $[\int_{\C_{1}} dt_1 ..., \int_{\C_{2}}
dt_2...]$, the result is independent of whether $\C_1$ lies inside
$\C_2$ or vice
versa.  The reason is that the change of variables $t_1 \rightarrow t_1^{-1}$,
$t_2 \rightarrow t_2^{-1}$ reverses inside and outside while leaving otherwise
equivalent expressions.  In our subsequent calculation we will always check
that commutators are the same computed both ways, since this provides a
powerful check on the algebra.

The equivalence of the contour integral (\ref{eq:deltan})
with the previous formulas for the hidden symmetry transformations
is demonstrated by deforming the contour ${\cal C}$ out to infinity
picking up the residues of any poles at $t = 0$ and $t = \infty$.  If $ n$
is a positive integer, the only pole is at $t = 0$, which obviously reproduces
the previous result $\Delta_n (\epsilon) = \delta_n (\epsilon)$.  If $-
n$ is a negative integer, the only pole is at $t = \infty$, which gives the
previous result $\Delta_{-n} (\epsilon) = \bar\delta_n (\epsilon)$.  Finally,
if $n = 0$ there are poles at both $t = 0$ and at $t = \infty$, so $\Delta_0
(\epsilon) = \delta_0 (\epsilon) + \bar\delta_0 (\epsilon)$.  This is a simpler
way to understand the results of Section 2.4.

The calculation of the algebra of these symmetries is remarkably easy:
\begin{equation}
[\Delta_{m} (\epsilon_1), \Delta_{n} (\epsilon_2)] g
= \int_{{\cal C}_{1}} {dt_1\over 2\pi i} t_1^{-m-1}
\int_{{\cal C}_{2}} {dt_2\over 2\pi
i} t_2^{-n -1} [\delta (\epsilon_1, t_1), \delta(\epsilon_2, t_2)]g,
\end{equation}
where we should deform the contours so that either ${\cal C}_1$
is inside ${\cal C}_2$ or
${\cal C}_2$ is inside ${\cal C}_1$.
Either choice will do. Now substitute the commutator
\begin{equation}
[\delta(\epsilon_1, t_1), \delta (\epsilon_2, t_2)] = {t_1 \delta
(\epsilon_{12}, t_1) - t_2 \delta (\epsilon_{12}, t_2)\over t_1 - t_2}
\end{equation}
and consider the two terms separately.
For the $\delta (\epsilon_{12}, t_1)$ term we can evaluate the $t_2$ integral
by shrinking the contour ${\cal C}_2$ to a point.  Similarly, for the $\delta
(\epsilon_{12}, t_2)$ term the $t_1$ integral can be evaluated by shrinking the
contour ${\cal C}_1$.
Whichever of the two contours started on the outside will give
a contribution from the residue of the pole at $t_1 = t_2$.  In either case,
the result is evidently
\begin{equation}
\int_{\cal C} {dt\over 2\pi i} t^{-m -n - 1} \delta (\epsilon_{12}, t)g
 = \Delta_{m + n} (\epsilon_{12})g.
\end{equation}
Thus we have a complete affine current algebra, without a central term:
\begin{equation}
[\Delta_{m} (\epsilon_1), \Delta_{n} (\epsilon_2)] =
\Delta_{m + n} (\epsilon_{12}) \quad m,\, n \in \ZZ.
\end{equation}
It is plausible, especially in view of the relationship to WZNW theory, that
the
algebra of the quantum theory has a central term.

It is now clear that $n$ is a discrete
momentum-like coordinate, so it is natural to
Fourier transform to an angular coordinate $\sigma$.  The charge densities
$J^i(\sigma) = \sum J_n^i e^{in\sigma}$
then satisfy the standard current algebra
\begin{equation}
[J^i (\sigma), J^j (\sigma')] = 2\pi f^{ij}{}_k J^k (\sigma)
\delta (\sigma-\sigma').
\end{equation}
The ``momentum'' $n$ and the ``position'' $\sigma$
should not be confused with the
physical momentum and position.  On the face of it, they are completely
unrelated.  Indeed, the physical coordinate $x^1$ describes an infinite line,
whereas $\sigma$ describes a circle. (Later, we will suggest that a doubled
algebra on a semicircle or line segment is a better interpretation.)

\sectiono{Virasoro Symmetry}

\subsection{The symmetry}

Having found current algebra symmetries in PCM's and SSM's, it is plausible
that they should also contain Virasoro symmetry algebras.  If the current
algebra symmetries were also present in the quantum theory, with central terms,
then the Sugawara construction would guarantee the existence of Virasoro
symmetries.  The trouble with this approach is that the Sugawara construction
is inherently quantum mechanical (the coefficient is proportional to
$\hbar^{-1}$), and we have only studied the classical theory.  As long as we
are unwilling to study the quantum theory, this means that other methods are
required to find the symmetry.  Of course, if we do find a Virasoro symmetry
in the classical theory, it should still be the
classical counterpart of the algebra that the Sugawara construction would
provide in the quantum theory.

Virasoro-like symmetries have been claimed by other authors for PCM's and
SSM's \cite{bib:cheng,bib:hou,bib:li}.
These works appear to utilize the transformation
\begin{equation}
\delta g = g \dot X (t) X (t)^{-1},
\end{equation}
where $\dot X (t) = {d\over dt} X(t)$.  Therefore, let us investigate what this
does to the PCM equation of motion $\partial \cdot A = 0$.  Using
\[ \delta A_\pm = D_\pm (\dot XX^{-1}) =
\partial_\pm (\dot XX^{-1}) + [A_\pm,
\dot XX^{-1}] \]
\begin{equation}
= - \dot\alpha_\pm A_\pm + (1 - \alpha_\pm) [ A_\pm, \dot XX^{-1}],
\end{equation}
and $\partial \cdot A = F_{+-} = 0$, as well as, $\alpha_\pm = t/(1 \mp t)$,
one obtains
\begin{equation}
\delta (\partial \cdot A) = {1\over t^2 - 1} [A_+, A_-].
\end{equation}
Thus, this is not a symmetry.  However, in the course of the calculation terms
of the form $[[A_+, A_-], \dot XX^{-1}]$ did cancel, which is encouraging.
This motivates us to look for another term that could give a compensating
contribution to $\delta (\partial \cdot A)$.

Let us now consider a transformation of the form $\delta^\I g = g \I$, where
\begin{equation}
\I = \int_{x_{0}}^x (A_+ dy^+ - A_- dy^-)
\end{equation}
is the contour-independent integral that first appeared in
eq. (\ref{eq:iappears}).  This
variation implies that
\begin{equation}
\delta^\I A_\pm = \pm A_\pm + [A_\pm, \I].
\end{equation}
{}From this it follows that
\begin{equation}
\delta^\I (\partial \cdot A) = - [A_+, A_-].
\end{equation}
Therefore, we deduce that the combined transformation
\begin{equation}
\delta^V (t) g = \lambda g (\dot X X^{-1} + {1\over t^2 - 1} \I)
\end{equation}
is a classical symmetry with a spectral parameter.  The coefficient $\lambda$
is an arbitrary scalar function of $t$, which will be chosen to have a
convenient form later. An alternative way of
understanding this result is to note that
\begin{equation}
\delta^V A_\pm = \pm {\lambda \over t}
\partial_\pm \left( \dot X X^{-1} +
{\I+t\J\over t^2 - 1}\right),  \label{eq:detavir}
\end{equation}
where $\J$ is the contour-independent expression
\begin{equation}
\J = \int_{x_{0}}^x ((2A_+ + [A_+, \I])dy^+ + (2A_- - [A_-, \I])dy^-).
\end{equation}
This makes it obvious that $\partial^\mu (\delta^V A_\mu) = 0$, and that
$\delta^V$ is therefore a classical symmetry.

The next step is to define $\delta_n^V g$ in analogy with
the contour integral representation of $\Delta_n
(\epsilon) g$ introduced in eq.(\ref{eq:deltan}):
\begin{equation}
\delta_n^V g = \int_\C {dt\over 2\pi i} t^{-n -1} \delta^V (t) g. = g \int_\C
{dt\over 2\pi i} t^{-n -1} \lambda (t) \left(\dot X X^{-1} + {\I\over t^2 -
1}\right).
\end{equation}
By the same reasoning as in Section 2.7, the change of variables $t \rightarrow
1/t$ can be used to restrict the possibilities for $\lambda (t)$.  Recalling
that
\begin{equation}
X (t^{-1}) = g^{-1} (x) \tilde{X} (t) g_0,
\end{equation}
which was derived in Section 2.4, we see that when $t \rightarrow t^{-1}$,
\begin{equation}
\dot XX^{-1} \rightarrow - t^2 g^{-1} (x) {\dot{\tilde X}} \tilde{X}^{-1} g
(x).
\end{equation}
Therefore, also using $g\I = - \tilde{\I} g$,
\begin{equation}
\delta_n^V g = \int_\C {dt\over 2\pi i} t^{n + 1} \lambda(t^{-1})
\left({\dot{\tilde X}} \tilde{X}^{-1} + {1\over t^2 - 1} \tilde{I} \right)g.
\end{equation}
What we want, in analogy with the current algebra case, is $\delta_n^V g =
\bar\delta_{-n}^V g$.  Since the infinitesimal parameter of a $\delta^V$
transformation is a scalar, rather than a matrix,
the present problem is simpler in one respect.  Specifically, whereas
$\bar \delta (\epsilon, t) = \tilde{\delta} (g_0 \epsilon g_0^{-1}, t)$ in the
current algebra case, $\bar\delta^V$ and $\tilde{\delta}^V$ are the same thing.
The transformation $\tilde{\delta}^V$ transforms $g^{-1}$ in the same way that
$\delta^V$ transforms $g$.  Thus,
\begin{equation}
\tilde{\delta}_n^V g = - g \int_\C {dt\over 2\pi i} t^{n + 1}
\lambda(t^{-1}) \left(\dot XX^{-1} + {1\over t^2 - 1} \I\right).
\end{equation}
Now, requiring that $\tilde{\delta}_{-n}^V = \delta_n^V$ gives the
condition $t \lambda (t^{-1}) = t^{-1} \lambda (t)$.  Therefore we set
\begin{equation}
\lambda (t) = (t^2 - 1) f (t + t^{-1}).
\end{equation}
Then
\begin{equation}
\delta_n^V g = g \int_\C {dt\over 2\pi i} t^{-n -1} f (t + t^{-1}) ((t^2 -
1) \dot XX^{-1} + \I)
\end{equation}
transforms correctly under $t \rightarrow 1/t$.  Moreover, as we
explained in Section 2.7, this will ensure that, when we study commutators, the
relative placement of contours will not matter.  It is a curious fact that if
$f(t + t^{-1})$ is regular at $t =  \pm1$, then the $\I$ contribution to this
integral vanishes.  Its presence or absence affects which contributions
arise from poles at $t=0$ and which arise from poles at $t=\infty$.

The conserved currents associated with the
hidden Virasoro symmetry of PCM's are
given by $D_\mu \Big((t^2 - 1) \dot X X^{-1} + \I \Big)$.
Therefore, the corresponding
conserved charges are
\[ Q^V (t) = \int_{-\infty}^\infty D_0
\Big((t^2 - 1) \dot X X^{-1} + \I\Big) dx^1 \]
\begin{equation}
= {1\over t} \int_{-\infty}^\infty \partial_1 \Big((t^2-1)
\dot X X^{-1} + \I + t \J \Big) dx^1
= {1 \over t}\Big((t^2 - 1) \dot X X^{-1} + \I
+ t \J \Big)\Big|_{-\infty}^{\infty}, \label{eq:pcmvircharge}
\end{equation}
where we have used eq.(\ref{eq:detavir}) with $\lambda = t^2 -1$
(corresponding to $f=1$).

\subsection{The algebra}

Having found the symmetry transformations, we can study the algebra generated
by $\Delta_n (\epsilon)$ and $\delta_n^V$.  This means studying the
commutators of
\begin{eqnarray}
\delta (\epsilon, t) g &=& g X \epsilon X^{-1}\\
\delta^V (t) g &=&  g ((t^2 - 1) \dot X X^{-1} + \I).
\end{eqnarray}
Where we have set $f(t + t^{-1})=1$.  The commutator
$[\delta (\epsilon_1, t_1), \delta (\epsilon_2, t_2)]$ was already studied in
Section 2.3,
so let us now consider $[\delta^V (t_1), \delta (\epsilon, t_2)] g =
[\delta_1^V, \delta_2] g.$  To begin with, this is
\begin{equation}
[\delta_1^V, \delta_2] g = \delta_1^V (g X_2 \epsilon X_2^{-1}) - \delta_2
(g ((t_1^2 - 1) \dot X_1 X_1^{-1} + \I)).
\end{equation}
The evaluation of this requires formulas for $\delta^V (t_1) X_2$ and
$\delta(\epsilon, t_2) \I$.  The transformation $\delta (\epsilon, t_2)X_1$ is
already given in eq. (\ref{eq:deltax}).

As usual, a convenient formula for $\delta^V (t_1) X_2 = \delta_1 X_2$ can be
obtained by solving differential equations obtained by varying the Lax pair
$(\partial_\pm + \alpha_{2\pm} A_\pm) X_2 = 0$.  The equations are
\begin{equation}
(\partial_\pm + \alpha_{2\pm} A_\pm) (\delta_1^V X_2)
+ \alpha_{2\pm} (\delta_1^V A_\pm ) X_2 = 0,
\end{equation}
where
\begin{equation}
\delta_1^V A_\pm =  (- \beta_1 \dot\alpha_{1\pm} \pm 1) A_\pm
+  [A_\pm, (1 - \alpha_{1\pm}) \beta_1 \dot X_1 X_1^{-1} + \I]
\end{equation}
and
\begin{equation}
\beta_1 = t_1^2 - 1.
\end{equation}
The easiest method of solution is to guess that the anwer has the structure.
\begin{equation}
\delta_1^V X_2 =  (\gamma_a \dot X_1 X_1^{-1} + \gamma_b \I
+ \gamma_c \dot X_2 X_2^{-1})X_2 \label{eq:dxvira}
\end{equation}
and derive equations for $\gamma_a, \gamma_b, \gamma_c$, which are scalar
functions that are allowed to depend on $t_1$ and $t_2$.  Note that all of
these terms vanish at $x_0$ so the boundary condition is already built in.

Substituting the ansatz for $\delta_1 X_2$ in the differential equations gives
the conditions
\begin{equation}
\alpha_{2\pm} (-\beta_1 \dot\alpha_{1\pm} \pm 1) - \dot\alpha_{1\pm} \gamma_a -
\dot\alpha_{2\pm} \gamma_c = 0
\end{equation}
\begin{equation}
\alpha_{2\pm} \gamma_b + \alpha_{2\pm} = 0
\end{equation}
\begin{equation}
(\alpha_{1\pm} - \alpha_{2\pm})\gamma_a - \alpha_{2\pm} (1 - \alpha_\pm)
\beta_1 = 0.
\end{equation}
The last two pairs of equations occur twice, so we actually are solving ten
equations for three unknowns.
Fortunately, there is a solution, and it is unique:
\begin{equation}
\gamma_a = {t_2 (t_1^2 -1)\over t_1 - t_2} ,\quad
\gamma_b = - 1, \quad \gamma_c = {t_1 (1
- t_2^2) \over t_1 - t_2}. \label{eq:dxvirb}
\end{equation}

Now, let us evaluate $\delta (\epsilon, t_2) \I = \delta_2 \I$.  Since we know
from Section 2.2 that
\begin{equation}
\delta_2 A_\pm = \pm t_2^{-1} \partial_\pm (X_2 \epsilon X_2^{-1}),
\end{equation}
we have
\begin{equation}
\delta_2 \I = {1\over t_2} \int_{x_{0}}^x \partial_\mu (X_2 \epsilon X_2^{-1})
dy^\mu = {1\over t_2} (X_2 \epsilon X_2^{-1} - \epsilon).
\end{equation}
One more useful preliminary is to use
\begin{equation}
\delta_2 X_1 = {t_1\over t_2 - t_1} (X_2 \epsilon X_2^{-1} X_1 - X_1 \epsilon),
\end{equation}
derived in Section 2.3, to deduce that
\begin{equation}\delta_2 (\dot X_1 X_1^{-1}) = {t_2\over (t_1 - t_2)^2} (X_2
\epsilon X_2^{-1} - X_1 \epsilon X_1^{-1}) + {t_1\over t_2 - t_1} [X_2 \epsilon
X_2^{-1}, \dot X_1 X_1^{-1}].
\end{equation}

Assembling the ingredients above gives
\[ [\delta_1^V, \delta_2] g =  \Big({1\over t_2}
(\delta (\epsilon, 0) - \delta (\epsilon, t_2)) +
{t_2 (t_1^2 - 1)\over (t_1 - t_2)^2} (\delta (\epsilon, t_1)
- \delta (\epsilon, t_2))\]
\begin{equation}
+ {t_1 (1 - t_2^2)\over t_1 - t_2} ~ {\partial\over\partial t_2} \delta
(\epsilon, t_2) \Big) g.  \label{eq:deltagpcm}
\end{equation}
So the algebra closes.  The next step is to compute
\begin{equation}
[\delta_{m}^V, \Delta_{n} (\epsilon)] g = \int_{\C_{1}} {dt_1\over
2\pi i} t_1^{-m - 1} \int_{\C_{2}} {dt_2\over 2\pi i} t_2^{-n -1}
[\delta_1^V, \delta_2]g.
\end{equation}
Integrating by parts the ${\partial\over\partial t_2} \delta (\epsilon, t_2)$
term and simplifying a bit gives
\[ [\delta_{m}^V, \Delta_{n} (\epsilon)] g = \int_{\C_{1}} {dt_1\over
2\pi i} t_1^{-n -1} \int_{\C_{2}} {dt_2\over 2\pi i} t_2^{-n -1}  \Bigg\{
{1\over t_2} \delta (\epsilon, 0)\]
\begin{equation}
+ {t_2 (t_1^2 - 1)\over (t_1 - t_2)^2} \delta (\epsilon, t_1) + n {t_1 (1 -
t_2^2)\over t_2 (t_1 - t_2)} \delta (\epsilon, t_2)\bigg\} g.
\end{equation}
The $\delta (\epsilon, 0)$ term vanishes upon doing the $t_2$ integration.

To make further progress we should stipulate how the contours are arranged.  If
$\C_2$ is outside $\C_1$ then the entire
answer comes from the $\delta (\epsilon, t_1)$ term:
\begin{equation}
[\delta_{m}^V, \Delta_{n} (\epsilon)] g = n \int_\C {dt\over
2\pi i} t^{-m -n -2} (t^2 -1) \delta (\epsilon, t) g.
\end{equation}
If, on the other hand, $\C_1$ is outside $\C_2$,
then the entire answer comes from the $\delta (\epsilon, t_2)$ term.  It is
precisely the same, as we built in when we restricted the form of $\lambda
(t_1)$.

Now let us reexpress the algebra representing $\Delta_{n} (\epsilon)$
and $\delta_{m}^V$ in terms of
generators $J_{n} (\epsilon) = J_{n}^i \epsilon^i$ and
$K_{m}$:
\begin{equation}
[K_{m}, J^i_{n}] = n (J^i_{m + n-1} - J^i_{m +
n + 1}). \label{eq:pcmalg}
\end{equation}
This is to be contrasted with what one would expect for Virasoro generators
$L_n$:
\begin{equation}
 [L_{m}, J^i_{n}] = - n J^i_{m + n}.
\end{equation}

The Jacobi identity for ($KJJ$) is easily checked.  The Jacobi identity for
($KKJ$) is
\begin{equation}
[[K_{m}, K_{n}], J^i_{p}] = p (m - n) (-J^i_{m
+ n + p -2} +2 J^i_{m + n + p} - J^i_{m +
n + p +2}].
\end{equation}
This allows us to infer that, up to terms that commute with $J_n$,
\begin{equation}
[K_{m}, K_{n}] = (m - n) (K_{m + n + 1} -
K_{m + n - 1}).
\end{equation}
This can be checked by commuting the appropriate contour integrals, which is
certainly a lot more tedious.  To make contact with the usual (centerless)
Virasoro algebra, suppose we identify
\begin{equation}
K_n = L_{n + 1} - L_{n - 1}.
\end{equation}
This certainly acounts for the algebra.  The more difficult question is whether
there is a realization of the $L_n$'s as symmetry operations of the theory.
If this is not possible then we simply have a subalgebra of Virasoro defined by
$K_n = L_{n + 1} - L_{n - 1}$.

\subsection{Non-extendability of the algebra}

The following discussion concerns the Virasoro algebra $\V$ and its
subalgebras.  Since the interest of this paper is in classical field theory the
algebras are assumed to have no central change.  However, the considerations
that follow could easily be generalized to include one.  Let us define
$\V_p \subset \V$ to be the subalgebra generated by
\begin{equation}
\L_n^{(p)} = L_n - L_{n+p}, ~~n \in \ZZ.
\end{equation}
In this notation the $K_n$'s of the preceding subsection are given by $K_n = -
\L_{n-1}^{(2)}$.  The algebra of $\V_p$ is
\begin{equation}
[\L_m^{(p)}, \L_n^{(p)}] = (m - n) (\L_{m+n}^{(p)} - \L_{m+n+p}^{(p)}).
\end{equation}

Now let us examine the conditions under which $\V$ can be reconstructed from
$\V_p$.  Toward this end let us define
\begin{equation}
\L_n^{(p)} (w) = \sum_{m=0}^\infty \L_{n+mp}^{(p)} w^m.
\end{equation}
Rearranging terms a little bit gives
\begin{equation}
\L_n^{(p)} (w) = L_n + (w - 1) \sum_{m=1}^\infty L_{n+mp} w^{m-1}.
\end{equation}
Therefore, the question is whether we can define $L_n$ in terms of the $\V_p$
generators by the formula
\begin{equation}
L_n = \lim_{w\rightarrow 1} \L_n^{(p)} (w) .
\end{equation}
Whether or not this makes sense depends on the particular realization of the
algebra under consideration.

To illustrate that this procedure makes sense, at least sometimes, let us
consider a realization of the classical Virasoro algebra in terms of a
stress tensor $T (z)$
\begin{equation}
L_n = \oint {dz\over 2\pi i} z^{n+1} T(z).
\end{equation}
Now suppose, we were only given the quantities
\begin{equation}
\L_n^{(p)} = \oint {dz\over 2\pi i} z^{n+1} (1 - z^p) T(z).
\end{equation}
Then, using the definition above
\begin{equation}
\L_n^{(p)} (w) = \oint {dz\over 2\pi i} z^{n+1} \left({1 - z^p\over 1 -
wz^p}\right) T (z).
\end{equation}
Clearly, the limit $w \rightarrow 1$ is well-defined in this case and gives the
desired result.

Now let us try to apply this procedure to the $\V_2$ symmetry algebra of
PCM's.  Using
\begin{equation}
K_n = \int_\C {dt\over 2\pi i} t^{-n -1} (t^2 - 1) \dot X (t) X (t)^{-1}
\end{equation}
and $\L_n^{(2)} = - K_{n + 1}$ gives
\begin{equation}
\L_n^{(2)} (x) = - \int_\C {dt\over 2\pi i} t^{-n} \left({t^2 - 1\over t^2 -
w}\right) \dot X (t) X (t)^{-1}.
\end{equation}
Now the crucial fact is that $\C = \C_+ + \C_-$, where
$\C_\pm$ are small circles
about $t = \pm 1$.  The poles at $t = \pm \sqrt{w}$ must be outside of these
contours.  Of course, the contour can then be deformed to have $\pm
\sqrt{w}$ on the inside if one compensates with the appropriate residues.
Calling the deformed contour $\C'$,
\[ \L_n^{(2)} (w) =  - \int_{\C'} {dt\over 2\pi i} t^{-n} \left({t^2 -
1\over t^2 - w}\right) \dot X (t) X (t)^{-1} \]
\begin{equation}
 + {1\over 2} (\sqrt{w})^{-n -1} (w - 1) \Big(\dot X (\sqrt{w}) X
(\sqrt{w})^{-1}\! - (-1)^n \dot X (-\sqrt{w}) X (-\sqrt{w})^{-1} \Big).
\end{equation}
Now, the limit $w \rightarrow 1$ can be considered.  The integral has a
well-defined limit.  However, the additional terms do not.  $\dot X (t) X
(t)^{-1}$ is more singular than $(t^2 - 1)^{-1}$ as $t \rightarrow \pm 1$.
Indeed, if this were not the case the integrals that define $K_n$ would vanish!
We conclude that $\V^{(2)}$ is the best we can do.  PCM's do not have a
complete Virasoro symmetry algebra, and there is no stress tensor
$T(z)$ associated to the hidden symmetry.
It would be interesting to know what
the implications of this fact are for the Sugawara construction in the
quantum theory.

Let us define $K(\sigma) = \sum K_n e^{in\sigma}$ and $T(\sigma) = \sum L_n
e^{in\sigma}$ in analogy with $J(\sigma)$, which we discussed in Section 2.7.
Then the relation $K_n = L_{n+1} - L_{n-1}$ implies that
\begin{equation}
T(\sigma) = {i\over 2} ~ {K(\sigma)\over\sin \sigma}.
\end{equation}
Since we actually only know the $K_n$'s and not the $L_n$'s, this can be
regarded as a definition of $T(\sigma)$.  The impossibility of reconstructing
the $L_n$'s from the $K_n$'s now has a simple explanation:  $K(\sigma)$ does
not vanish at $\sigma = 0, \pi$.  This result is surprising inasmuch as the
study of the current algebra symmetry showed perfect rotational symmetry in
$\sigma$.  Now we are discovering that $\sigma = 0, \pi$ are actually special
points, so this rotational symmetry is not a general feature of the theory.
When we study symmetric space models in Section 5 we will discover that the
points $\sigma = 0$ and $\sigma = \pi$ have a special status already for
the current algebra in that case. By decomposing the Fourier series
expansions of $K$, $T$, and the currents $J^i$  into sines and cosines,
in the usual way, one can replace each of them by a pair of currents
on the interval $0 \leq \sigma \leq \pi$, one of which satisfies
Neumann boundary conditions and one of which satisfies Dirichlet
boundary conditions at the ends. Doing this, one sees that the
singular behavior at $\sigma = 0, \, \pi$ resides entirely in the
Dirichlet part of $T$, which corresponds to the Neumann part of $K$.

\sectiono{Non-Abelian Duality Transformation}

\subsection{A change of variables}

In this subsection, we describe a rewriting of the 2D PCM that results from a
certain non-linear and non-local change of variables, which we interpret
as a non-abelian duality transformation.
Two different non-abelian duality transformations have been considered
previously in the
literature, both of which introduce a Lie-algebra valued field $\phi (x)$
that
is ``dual'' to $g (x)$.  The first, given by $A_\pm = \pm \partial_\pm \phi$,
implements $\partial \cdot A = 0$ as a Bianchi identity, and satisfies an
equation of motion determined by $F_{+-} = 0$.  Nappi has shown that the
theory
expressed in terms of $\phi$, though classically equivalent to the PCM, is
inequivalent at the quantum level \cite{bib:nappi}.  An alternative duality
transformation is given by $A_\pm = \pm D_\pm \phi$, which is also consistent
with $\partial \cdot A = F_{+-} = 0$.  Fridling and Jevicki studied this (for
the special case of SU(2)) and showed
(using path integral methods) that it is quantum mechanically
equivalent to the original PCM \cite{bib:fridling}.
This was further elaborated upon by Fradkin and Tseytlin \cite{bib:fradkin}.

The transformation we will consider differs from
the previous ones in two important respects.  (Its quantum properties
are beyond the scope of this paper.)
First, the transformation contains
a free parameter.  Second, the dual field is group valued
like the original field (rather than
Lie-algebra valued). The change of variables is motivated by
the structure of the hidden symmetry $\delta g = g X \epsilon X^{-1}$.
This suggests considering a new group-valued variable
\begin{equation}
g' = g X (u) , \quad - 1< u < 1.
\end{equation}
The transformation is parametrized by $u$ (and an implicit base point $x_0$).
$X(u)$ is very singular for $u = \pm 1$, so it is important that those values
are excluded.  Restriction to the interval $-1< u < 1$ ensures
that there is a convergent series expansion about $u = 0$, though this fact
will not be utilized.

The classical equation of motion that $g'$ satisfies is easy to deduce.  Using
$\partial_\pm X = - \alpha_\pm (u) A_\pm X$, one finds that
\begin{equation}
\tilde{A}'_\pm = (1 - \alpha_\pm (u)) \tilde{A}_\pm,
\end{equation}
where, as before,
\begin{equation}
\tilde{A}_\pm = g \partial_\pm g^{-1} \quad {\rm and} \quad \tilde{A}'_\pm = g'
\partial_\pm g^{\prime-1}.
\end{equation}
Substituting in the equation of motion $\partial_+ \tilde{A}_- +\partial_-
\tilde{A}_+ = 0$, and using $\alpha_\pm (u)= u/( u\mp 1)$, gives
\begin{equation}
(1 + u) \partial_+ \tilde{A}'_- + (1 - u) \partial_- \tilde{A}'_+ = 0.
\end{equation}
Equivalently, defining  ${A}'_\pm = g^{\prime-1}
\partial_\pm g'$, one can write
\begin{equation}
(1 - u) \partial_+ A'_- + (1 + u) \partial_- A'_+ = 0. \label{eq:wzeqn}
\end{equation}
The interesting fact is that this equation of motion is obtained from the
action
\begin{equation}
S_u (g') = S_{PCM} (g') + u S_{WZ} (g'),
\end{equation}
where $S_{WZ}$ denotes the standard Wess--Zumino term.  The values $u = \pm 1$,
which we are not allowed to use, correspond to
Wess--Zumino--Novikov--Witten (WZNW)
theory.  This fact is evident from the structure of the equations of motion
given above.  Thus, we have shown that for all values of $u$,
other than $\pm 1$,
$S_u$ describes the same classical theory as $S_0 = S_{PCM}$.
The addition of a WZ term has been interpreted as introducing torsion
on the group manifold in ref. \cite{bib:braaten}.

In order to better understand the significance of this change of variables, it
is instructive to consider a trivial free scalar field theory with ${\cal L} =
\partial_+ \varphi \partial_- \varphi$.
Identifying $g = \exp \, \varphi$ and writing a
general classical solution as $\varphi^+ (x^+) + \varphi^- (x^-)$,
one sees that a
$u$ transformation corresponds to
\begin{equation}
\varphi' = (1 - \alpha_+) \varphi^+ + (1 - \alpha_-) \varphi^- + ~{\rm const}.
= {1\over 1 - u^2} (\varphi - u \tilde{\varphi}) + ~{\rm const}.\ ,
\end{equation}
where $\tilde{\varphi} = \varphi^+ - \varphi^-$.
Thus a $u$ transformation mixes
$\varphi$ with its dual $\tilde\varphi$,
and a complete interchange corresponds to $u \rightarrow \infty$.
Rescaling by a factor of $1 - u^2$, one sees that the theory at $u = \pm 1$
describes a chiral boson.  Therefore, in a sense, one could say that the theory
of a free chiral boson is an abelian WZNW theory.  Of course, the Wess--Zumino
term vanishes in this case.  The conclusion, in general, is that a $u$
transformation can be regarded as implementing a non-abelian duality
transformation.

Even though the only purpose of this paper is to analyze classical theories, we
would be remiss not to insert a comment about quantum physics at this
point.  As is well-known \cite{bib:wittenc},
in the quantum theory the coefficient of the
Wess--Zumino term should be an integer $k$.  The parameter $u$ that we have
introduced is the {\it ratio} of the coefficients of the PCM term and the WZ
term.  It is this ratio that is $\pm1$ for WZNW theory
\cite{bib:wess,bib:novikov,bib:wittenb}.  So, in general, for the quantum
theory
\begin{equation}
S_{k,u}= k\Big( {1 \over u} S_{PCM} + S_{WZ}\Big).
\end{equation}
Based on our experience with WZNW theory, it is
plausible that the quantum theory does depend on the choice of $k$.  Moreover,
it would not be surprising to discover that the quantum-mechanical central
extension of the infinite-dimensional symmetry algebras $\hat G$ and $\V_2$ are
controlled by $k$.  A possible approach to thinking about these issues, which
has been very fruitful for WZNW theory \cite{bib:wittenb}, would be to seek
fermionic formulations that are equivalent at the quantum level. Some
possibilities along these lines have been discussed in the literature
\cite{bib:polyakovc,bib:rajeev}.

\subsection{Current algebra symmetry of the $u$-transformed theory}

Rather than tracing through what the hidden symmetry of the $u = 0$ theory
implies for the $u$-transformed theory, let us discover the hidden symmetry of
the transformed theory directly by generalizing the
discussion of Section 2.\footnote{This has been done previously
(in the supersymmetric case!) by
Chau and Yen \cite{bib:chauc}.}
Now let's take $u$ to have a fixed value
and introduce a spectral parameter $t$, just as we did in the
case of $u = 0$.  From now on, the prime is dropped in the $u$-transformed
theory.  As before, we begin by determining a Lax pair
\begin{equation}
(\partial_\pm + \alpha_\pm (t, u) A_\pm) X(t,u) = 0,
\end{equation}
where $\alpha_\pm (t,u)$ and $X(t,u)$ should reduce to the
expressions given in Section 2.2
when we set $u = 0$.  Using the equation of motion (\ref{eq:wzeqn})
and the Bianchi identity $F_{+ -} =0$, it is
easy to see that the compatibility condition for the Lax pair is
\begin{equation}
{1 + u\over\alpha_+} + {1 - u\over\alpha_-} = 2.
\end{equation}
There are many different ways to parametrize solutions to this equation.  The
one that will prove most convenient for our purposes is
\begin{equation}
\alpha_\pm (t,u) = {t\over t - \sigma_\pm (u)},
\end{equation}
\begin{equation}
\sigma_+ (u) = \sqrt{{1-u \over 1 + u}},
\quad \sigma_- (u) =  - \sqrt{{1 + u\over 1
- u}}.
\end{equation}
The branches are unambiguous for the range of values $- 1< u < 1$ that we
allow. As before, we set
\begin{equation}
\delta (\epsilon, t) g = g X (t,u) \epsilon X(t,u)^{-1},
\end{equation}
where
\begin{equation}
X (t,u) = P \exp \{ - \int_{x_{0}}^x (\alpha_+ A_+ dy^+ + \alpha_- A_- dy^-) \}
\end{equation}
is contour independent.  These are the same formulas as before except that $u$
dependence is now introduced via the quantities $\alpha_\pm (t,u)$.  Note that
$X(t,u)$ has a convergent power series expansion about $t = 0$ for $|t| < \min
\left(\sqrt{{1-u\over 1 + u}}, \sqrt{{1 + u\over 1 - u}}\right)$.  This circle
shrinks to a point as $u \rightarrow \pm 1$, corresponding to WZNW
theory.

Next, we analyze the commutator $[\delta (\epsilon_1, t_1), \delta (\epsilon_2,
t_2)]$ by the same procedure as before.  Let us again use the short-hand
notation $\alpha_{i\pm} = \alpha_\pm (t_i, u), X_i = X(t_i, u),$ etc.  Then
using
\begin{equation}
\delta_1 A_\pm = (1 - \alpha_{1\pm}) [A_\pm, \eta_1]
\end{equation}
and varying the Lax pair one can derive
\begin{equation}
(\partial_\pm + \alpha_{2\pm} A_\pm) \delta_1 X_2 = \alpha_{2\pm}
(\alpha_{1\pm} - 1) [A_\pm, \eta_1] X_2
\end{equation}
and
\begin{equation}
(\partial_\pm + \alpha_{2\pm} A_\pm) (\eta_1 X_2 - X_2 \epsilon_1) =
(\alpha_{2\pm} - \alpha_{1\pm}) [A_\pm, \eta_1]X_2.
\end{equation}
Now comes the step that depends on the specific choice of parametrization of
$\alpha_\pm (t,u)$.  With the choice given, one has
\begin{equation}
\alpha_{2\pm} (\alpha_{1\pm} - 1) = {t_2\over t_1 - t_2} (\alpha_{2\pm} -
\alpha_{1\pm}).
\end{equation}
This enables us to conclude, just as we did in the $u = 0$ case, that
\begin{equation}
\delta_1 X_2 = {t_2\over t_1 - t_2} (\eta_1 X_1 - X_2 \epsilon_1).
\end{equation}
Both sides of this equation satisfy the same differential equations and
boundary conditions (at $x = x_0$), which implies the equality.  It now
follows, step-by-step as before, that
\begin{equation}
[\delta (\epsilon_1, t_1), \delta (\epsilon_2, t_2)]g = {t_1 \delta
(\epsilon_{12},
t_1) - t_2 \delta (\epsilon_{12}, t_2)\over t_1 - t_2}g.
\end{equation}
The additional transformation $\bar\delta (\epsilon, t)$ can also be
introduced as before.

One shouldn't be surprised, perhaps, that the $u$-transformed theory has the
same symmetry algebra as the untransformed one.  There is one aspect of the
result that is a bit surprising, however.  The realization of the
affine current algebra symmetry
obtained this way is not the same as the one obtained by applying the $u$
transformation to the symmetry of the $u = 0$ theory.

\sectiono{Symmetric Space Models}

\subsection{Basic facts and formalism}

Let $G$ be a simple Lie group and $H$ a subgroup.  Then the Lie algebra ${\cal
G}$ can be decomposed into the Lie algebra ${\cal H}$ and its orthogonal
complement ${\cal K}$, which
contains the generators of the coset  $G/H$.  $G/H$ is
called a symmetric space if $[{\cal K, K}] \subset {\cal H}$ -- that is, if the
commutator of any pair of coset generators belongs
to the subalgebra ${\cal H}$.
In this section we wish to describe symmetric space models (SSM) analogous to
the PCM's of the preceding sections.  One point that should be noted right away
is that when $G$ is simple
PCM's and SSM's are distinct.  Neither is a special case of the other,
and there are no non-abelian theories that belong to both categories.
Later, we will discuss how to use a non-simple group
to formulate PCM's as SSM's. This will have interesting consequences.

Let us now restrict attention to the case of a simple group $G$.
For a physically acceptable theory
(with positive kinetic energy), the generators in $\K$ should all
be hermitian (non-compact case) or antihermitian (compact case). This allows
two classes of SSM's -- compact and noncompact.  The ones that occur
in supergravity and superstring theories are always noncompact.  A non-compact
SSM is very easy to describe, since there is a
unique symmetric space associated to $G$ that is physically acceptable.
It is given by $G/H$, where $H$ is
the maximal compact subgroup of $G$.  The proof is trivial.  The generators of
$H$ are antihermitian and those of $\K$ are hermitian.  Therefore, since
the commutator of two hermitian matrices is antihermitian,
$[{\cal K, K}] \subset {\cal H}$.  Compact symmetric spaces can be put
in one-to-one correspondence with these
non-compact ones.  They are given by the anti-hermitian part of the
complexification of $\G$.
In other words, all the generators of ${\cal K}$ are
multiplied by a factor of $i$.  By far the most commonly studied SSM's are the
compact ones based on $S^n = O(n+1)/O(n)$ and $CP^n = U (n+1)/U(n) \times
U(1)$.  The former is usually called ``the nonlinear sigma model.''  The
corresponding non-compact symmetric spaces are $O(n,1)/O(n)$ and $U(n,1)/ U(n)
\times U(1)$.  Much of what is known about the subject has been learned
from these examples.  The discussion that follows applies to an arbitrary
non-compact symmetric space based on a simple group $G$.
Later, we will indicate what changes are required
for the compact case and discuss a class of examples
based on groups that are not simple.

Let $g(x)$ be an arbitrary $G$-valued field, just as for a PCM.  To construct
an SSM, we associate local $H$ symmetry with left multiplication and global $G$
symmetry with right multiplication.  Thus, we require invariance under
infinitesimal transformations of the form
\begin{equation}
\delta g = - h (x) g + g \epsilon \quad h\in {\cal H}, \  \epsilon\in {\cal G}.
\end{equation}
It is evident that
\begin{equation}
g \partial_\mu g^{-1} = P_\mu + Q_\mu
\end{equation}
is $G$ invariant and Lie-algebra valued.  $P_\mu$, the hermitian part, is
${\cal K}$-valued and $Q_\mu$, the anti-hermitian part, is ${\cal H}$-valued.
Defining ${\cal D}_\mu g = (\partial_\mu + Q_\mu) g$, we have
\begin{equation}
({\cal D}_\mu + P_\mu) g = 0.
\end{equation}
{}From this we can deduce the Bianchi identity
$ [{\cal D}_\mu + {P}_\mu, {\cal D}_\nu + P_\nu] = 0 $.
Separating ${\cal H}$ and ${\cal K}$ components gives the well-known formulas
\begin{equation}
\D_\mu P_\nu - \D_\nu P_\mu = 0
\end{equation}
\begin{equation}
\partial_\mu Q_\nu - \partial_\nu Q_\mu + [Q_\mu, Q_\nu] +[P_\mu, P_\nu] = 0.
\end{equation}

As we have already noted, $P_\mu$ and $Q_\mu$ are invariant under global $G$
transformations.  Under local $H$ transformations $(\delta g = - hg)$
\begin{equation}
\delta Q_\mu = \D_\mu h ~{\rm and}~ \delta P_\mu = [P_\mu, h].
\end{equation}
Thus $Q_\mu$ transforms like an $H$ gauge field.  Now let us define
\begin{equation}
A_\mu = - 2 g^{-1} P_\mu g,
\end{equation}
which is evidently invariant under local $H$ transformations.  Note that $g$,
like a vielbein in general relativity,
is used to convert a local $H$ tensor to a
global $G$ tensor (or vice-versa).  Using
\begin{equation}
P_\mu = {1\over 2} (g \partial_\mu g^{-1} + \partial_\mu g^{-1\dagger}
g^{\dagger}),
\end{equation}
it is easy to show that
\begin{equation}
A_\mu = M^{-1} \partial_\mu M, \label{eq:aformula}
\end{equation}
where
\begin{equation}
M = g^\dagger g.
\end{equation}
$M$ parametrizes
the symmetric space $G/H$
without extra degrees of freedom.
The construction is analogous to forming the metric tensor
out of the vielbein in general relativity.  Since $A_\mu$ is written in a form
that is pure gauge, it is obvious that
\begin{equation}
F_{\mu\nu} = \partial_\mu A_\nu - \partial_\nu A_\mu + [A_\mu, A_\nu] = 0.
\end{equation}

Let us now define a classical field theory in flat space-time based on the
non-compact symmetric space $G/H$ by
\begin{equation}
\L = \eta^{\mu\nu} tr (A_\mu A_\nu) = 4 \eta^{\mu\nu} tr (P_\mu P_\nu),
\end{equation}
and vary this to find the equation of motion.  Under an arbitrary
infinitesmal variation $g^{-1} \delta g = \eta (x) \in \G$ we have
\begin{equation}
\delta M = \eta^\dagger M + M \eta,
\end{equation}
which implies that
\begin{equation}
\delta A_\mu = D_\mu (\eta + M^{-1} \eta^\dagger M).
\label{eq:deltaamu}
\end{equation}
Note that we define $D = \partial + A$, whereas $\D = \partial + Q$.
We now have
\begin{equation}
\delta \L = 2\, \tr [A^\mu D_\mu (\eta + M^{-1} \eta^\dagger M)]
= 2\, \tr  [ A^\mu \partial_\mu (\eta + M^{-1} \eta^\dagger M)].
\end{equation}
This unambiguously implies that  the equation of motion is
$\partial_\mu A^\mu = 0$.  In terms of $M$ this becomes
\begin{equation}
\partial_\mu \partial^\mu M = (\partial^\mu M)M^{-1} (\partial_\mu M).
\end{equation}

\subsection{Hidden symmetry}

Now we are ready to restrict attention two dimensions and look for the hidden
symmetry of SSM's.  Since we have $\partial \cdot A = 0$ and $F_{+-} = 0$, just
as for PCM's, we can once again form the Lax pair
\begin{equation}
(\partial_\pm + \alpha_\pm A_\pm) X = 0,
\end{equation}
with
\begin{equation}
\alpha_\pm (t) = {t\over t \mp 1}
\end{equation}
and solve for $X$:
\begin{equation}
X(t) = P \exp \Big(- \int_{x_{0}}^x (\alpha_+ A_+ dy^+
+\alpha_- A_- dy^-) \Big).
\end{equation}
The obvious guess now is that, just as for PCM's, the hidden symmetry is
described by $\delta g = g\eta$ with $\eta = X(t) \epsilon X(t)^{-1}$, where
$\epsilon$ is an arbitrary infinitesimal
constant element of the Lie algebra $\G$, that is
\begin{equation}
\delta g = g X(t) \epsilon X(t)^{-1} . \label{eq:casym}
\end{equation}

To test the conjectured symmetry, we must examine whether
$\delta (\partial \cdot
A) = 0$.  Using eq. (\ref{eq:deltaamu})
\begin{equation}
\delta (\partial \cdot A) = \partial \cdot D \eta + \partial \cdot D(M^{-1}
\eta^\dagger M).
\end{equation}
By comparison with the PCM case, we know that $\partial \cdot D\eta$
vanishes, so we need to examine the second term.  It may be rewritten as
\begin{equation}
\partial \cdot D (M^{-1} \eta^\dagger M) =
\partial^\mu (M^{-1} \partial_\mu \eta^\dagger M).
\end{equation}
Substituting
\begin{equation}
\partial_\pm \eta^\dagger = \alpha_\pm [A_\pm^\dagger, \eta^\dagger],
\end{equation}
and using $\partial \cdot A = 0$ and
$\alpha_+ + \alpha_- = 2 \alpha_+ \alpha_-$,
it is easy to see that this vanishes. Alternatively,
one can note that
\begin{equation}
\delta A_{\pm} = D_{\pm} (\eta + M^{-1} \eta^{\dagger} M) = \pm \partial_{\pm}
(t^{-1} \eta + t M^{-1} \eta^{\dagger} M), \label{eq:detassm}
\end{equation}
and so $\delta \partial \cdot A = 0$
vanishes.  Thus, we have indeed found the desired symmetry.

The conserved currents associated with the hidden current algebra symmetry of
SSM's are given by $D_\mu (\eta + M^{-1} \eta^{\dagger} M)$.  Therefore, the
corresponding conserved charges are
\[ Q (\epsilon, t) = \int_{-\infty}^\infty D_0 (\eta + M^{-1}
\eta^{\dagger} M)dx^1 \]
\begin{equation}
 =  \int_{-\infty}^{\infty} \partial_1 \Big({1\over t} \eta +
t M^{-1} \eta^{\dagger} M \Big) dx^1
= \left({1\over t} \eta +
t M^{-1} \eta^{\dagger} M\right)\Big|_{-\infty}^\infty,
\end{equation}
where we have used eq.(\ref{eq:detassm}).

Next we should examine the algebra $[\delta_1, \delta_2]g$, where $\delta_i =
\delta (\epsilon_i, t_i)$.  We know from the PCM analysis that the key step is
to find a useful expression for $\delta_1 X_2$ by varying $(\partial_\pm +
\alpha_{2\pm} A_\pm) X_2 = 0$.  This gives the differential equations
\begin{equation}
(\partial_\pm + \alpha_{2\pm} A_\pm) \delta_1 X_2 + \alpha_{2\pm} D_\pm (\eta_1
+ M^{-1} \eta_1^\dagger M)X_2 = 0.
\end{equation}
As before, we seek a solution of these equations satisfying the boundary
condition $\delta_1 X_2|_{x_{0}} = 0$.  Conjecturing that the extra term
has the form
\begin{equation}
(\delta_1 X_2)_{\rm extra} = \lambda(M^{-1} \eta_1^\dagger MX_{2}
-X_2 M_0^{-1}\epsilon_1^{\dagger} M_0)
\end{equation}
leads to the conditions
\begin{equation}
(\lambda +\alpha_{2\pm})\alpha_{1\pm} + \lambda (\alpha_{2\pm} - 1) =0,
\end{equation}
which is solved by
\begin{equation}
\lambda = {t_1 t_2 \over 1 - t_1 t_2} \label{eq:lambdaeq}
\end{equation}
Therefore the unique solution is
\begin{equation}
\delta_1 X_2 = {t_2\over t_1 - t_2} (\eta_1 X_2 - X_2 \epsilon_1) +
{t_1 t_2\over 1- t_1 t_2} (M^{-1} \eta_1^\dagger MX_{2}
-X_2 M_0^{-1}\epsilon_1^{\dagger} M_0), \label{eq:dxssm}
\end{equation}
where $M_0 = M(x_0)$.
The first term is the one we found for PCM's.  The second term, which is new,
is required to compensate
for the extra piece of $\delta A_\mu$ that occurs for
SSM's.

Now we can analyze the algebra by substituting $\delta_1 \eta_2 = [\delta_1 X_2
\cdot X_2^{-1}, \eta_2]$ into
\begin{equation}
[\delta_1, \delta_2] g = g ([\eta_1, \eta_2] + \delta_1 \eta_2 - \delta_2
\eta_1).
\end{equation}
This gives the result
\begin{equation}
[\delta_1, \delta_2] g = g \left({t_1 \eta (\epsilon_{12}, t_1) - t_2 \eta
(\epsilon_{12}, t_2)\over t_1 - t_2}\right) + \delta' g + \delta'' g,
\end{equation}
where
\begin{equation} \delta' g = {t_1 t_2\over 1 - t_1 t_2 } g ([M^{-1}
\eta_1^\dagger
M, \eta_2] - [ M^{-1} \eta_2^\dagger M, \eta_1]),
\end{equation}
\begin{equation} \delta'' g = {t_1 t_2\over 1- t_1 t_2 } g
(X_2 \epsilon'_{12} X_2^{-1} - X_1 \epsilon'_{21}X_1^{-1}),
\end{equation}
and
\begin{equation}
\epsilon'_{12} = M_0^{-1} \epsilon_1^{\dagger} M_0 \epsilon_2
-\epsilon_2 M_0^{-1}\epsilon_1^{\dagger} M_0.  \label{eq:parameter}
\end{equation}

The first term is the PCM result, so we need to interpret $\delta' g$
and $\delta'' g$.  The key step is to rewrite $\delta' g$
in the form $\delta' g = - h_{12} (x) g$, where
\begin{equation}
h_{12} (x) = (g^\dagger)^{-1} (\eta_1^\dagger M \eta_2 - \eta_2^\dagger M
\eta_1) g^{-1}
+ g (\eta_1 M^{-1} \eta_2^\dagger - \eta_2 M^{-1} \eta_1^\dagger) g^\dagger.
\end{equation}
Since $h_{12}$ is an anti-hermitian element of the Lie algebra,
$h_{12} \in
\H$.  This means that $\delta' g$ is a local $H$ transformation.  In
particular, since $M$ is invariant under local $H$ transformations,
$\delta' M =
0$.  Thus, restricted to $M$, the symmetry algebra is the pure current
algebra result plus the $\delta''$ contribution. Specifically,
\begin{equation}
[\delta_1, \delta_2]M = {t_1 \delta (\epsilon_{12}, t_1) -
t_2 \delta (\epsilon_{12}, t_2)\over t_1 - t_2} M + {t_1 t_2\over 1 - t_1 t_2 }
\left( \delta(\epsilon'_{12}, t_2) - \delta(\epsilon'_{21},t_1)\right) M.
\label{eq:ssmalgebra}
\end{equation}
The second term is a bit of a surprise, since it is frequently asserted
that the symmetry of SSM's is $\hat G$. However, this is not precisely correct.
The extra term is very important, and must not be dropped. In section 6.4
we will argue that without the extra term we would be confronted with
paradoxes. The result obtained here has been found previously for the
particular case
of the $S^n$ non-linear sigma models by Wu and also by Jacques and Saint-Aubin
\cite{bib:wu}.

\subsection{Compact symmetric space models}

The analysis in the preceding sections applies to non-compact SSM's.  As we
have
said, for simple groups $G$
each compact SSM is the partner of a non-compact one.  In the compact
case, the matrices $g(x)$ are unitary, and so the formula $M = g^\dagger g$ is
certainly not appropriate.  Fortunately, this formula can be rewritten in a
form that does generalize suitably.  The key fact is that symmetric spaces
have an involutive automorphism. This means that to each non-compact
Lie group $G$ with maximal compact subgroup $H$
we can associate a constant matrix $L$, satisfying $L^2 = 1$,
such that
$L$ commutes with the elements of $\H$ and anticommutes with elements
of $\K$. Now let us define
\begin{equation}
\tilde{g} = L g^{-1} L.
\end{equation}
This is the same as $g^\dagger$ in the non-compact case, but when we pass to
the corresponding compact group (by multiplying the coset generators by $i$) it
differs from $g^\dagger$.  Therefore, the definition
\begin{equation}
M = \tilde{g} g
\end{equation}
agrees with the previous formula in the non-compact case and is a plausible
formula in the compact case, too.  $M$ defined this way is an element of $G$
satisfying the restriction $\tilde{M} =  M$. This is just what is required
to represent a general element of the coset manifold.

Let us illustrate what has just been
said with a couple of examples.
The noncompact group $O(m,n)$ has maximal compact subgroup $O(m) \times
O(n)$.  In this example the matrix $L$ is
\begin{equation}
L = \left(\begin{array}{cc} 1_m & 0\\ 0 & -1_n \end{array}\right),
\end{equation}
where $1_m$ denotes an $m \times m$ unit matrix.
For the corresponding compact SSM, $O(m+n)/O(m) \times O
(n)$, the matrix $M$ is an $O(m+n)$ element that is subject to the restriction
$M = \tilde{M}$, which implies that $ML$ is symmetric.  A second class of
non-compact symmetric spaces is described by $Sp (2n, I\!\!R)/U(n)$.  In this
case the corresponding compact space $USp(2n)/U(n)$ is described using
\begin{equation}
L = i\left(\begin{array}{cc} 0 & 1_n\\ -1_n & 0 \end{array} \right).
\end{equation}
This time $\tilde{M} = M$ implies that $ML$ is hermitian. For $n=1$ these
are the same as $SL(2, \RR)/SO(2)$ and $SU(2)/U(1)$, respectively.

Because of the relationship between non-compact and compact SSM's described
above, the derivation of the hidden symmetry and its algebra is easily extended
to the compact case.  All that is required is to make the replacements
$g^\dagger \rightarrow \tilde{g}$ and $\eta^\dagger \rightarrow \tilde{\eta} =
- L\eta L$ in the formulas of the preceding subsection.

\subsection{Principal chiral models as symmetric space models}

Chau and Hou have pointed out that a PCM with compact Lie group $G$ can be
recast as a $(G \times G)/G$ SSM \cite{bib:chau}.
Their argument goes as follows.  Let us
write an arbitrary element of $G \times G$ in the form
\begin{equation}
g_S (x) = \left(\begin{array}{cc} g_L (x) & 0 \\ 0 & g_R (x)\end{array}
\right),
\end{equation}
and consider the SSM that corresponds to the involutive automorphism
\begin{equation}
L = \left(\begin{array}{cc}0 & 1\\ 1 & 0\end{array}\right).
\end{equation}
Then, using the formulas of the preceding subsection,
\begin{equation}
\tilde{g}_S = Lg_S^{-1} L = \left(\begin{array}{cc} g_R^{-1} & 0 \\
0 & g_L^{-1}\end{array}\right),
\end{equation}
and
\begin{equation}
M = \tilde{g}_S g_S = \left(\begin{array}{cc}  g_R^{-1}g_L & 0 \\ 0 &
g_L^{-1}g_R\end{array} \right) .
\end{equation}
The subgroup consists of ``diagonal'' elements for which $g_L = g_R$.  Next
define $g (x) =  g_R (x)^{-1}g_L (x)$, so that
\begin{equation}
M(x) = \left(\begin{array}{cc} g(x) & 0 \\ 0 & g(x)^{-1} \end{array} \right).
\end{equation}
Then
\begin{equation}
A_{S\mu} = M^{-1} \partial_\mu M = \left(\begin{array}{cc} g^{-1} \partial_\mu
g & 0 \\ 0 & g \partial_\mu g^{-1} \end{array}\right) = \left(\begin{array}{cc}
A_\mu & 0 \\ 0 & \tilde{A}_\mu \end{array} \right).
\end{equation}
The notation consists of attaching a subscript $S$ to all matrices that are $2
\times 2$ blocks (except for $M$ and $L$).  We now have
\begin{equation}
\L = \tr (A_S^\mu A_{S\mu} ) = \tr (A^\mu A_\mu) + \tr (\tilde{A}^\mu
\tilde{A}_\mu) = 2 \tr (A^\mu A_\mu),
\end{equation}
which describes the PCM for the group $G$.

This reformulation of a PCM as an SSM provides a new
perspective on the symmetry algebra of SSM's obtained in
the preceding subsection.  If the result there had been an affine current
algebra, without the extra piece that we found, then this reformulation of the
PCM would imply that it has $\hat G \times \hat G$ symmetry -- like
a WZNW theory!  We know that a PCM has one $\hat G$ algebra, but it certainly
does not have two, so it is fortunate that the symmetry algebra of an SSM is
not a usual affine algebra.  To make an iron-clad case, we now propose to show
that the peculiar symmetry algebra of SSM's that we have found
is exactly what is required in
order to reproduce the known (non-controversial) symmetry of PCM's.

The symmetry of the PCM, formulated as an SSM, is given by
\begin{equation}
\delta_S (\epsilon_S, t) M = M\eta_S + \tilde{\eta}_S M,
\end{equation}
where
\begin{equation}
\eta_S = X_S (t) \epsilon_S X_S (t)^{-1} \quad {\rm and} \quad
\tilde{\eta}_S = - L \eta_S L
\end{equation}
\begin{equation}
\epsilon_S = \left(\begin{array}{cc} \epsilon_L & 0\\ 0 &
\epsilon_R\end{array}\right)
\end{equation}
\begin{equation}
X_S (t) = P \exp (- \int_{x_{0}}^x (\alpha_+ A_{S+} dy^+ + \alpha_- A_{S-}
dy^-)).
\end{equation}
All these formulas are block diagonal.  Moreover, the upper and lower blocks of
$\delta_S M$ give equivalent formulas, namely
\begin{equation}
\delta_S g = g X \epsilon_L X^{-1} - \tilde{X} \epsilon_R \tilde{X}^{-1} g
= \big(\delta (\epsilon_L, t) + \tilde{\delta} (\epsilon_R, t)\big) g,
\label{eq:deltas}
\end{equation}
where we have recognized the transformations $\delta$ and $\tilde{\delta}$
defined in Section 2.  The question now is whether
\begin{equation}
[\delta_{S1}, \delta_{S2}] g = [\delta_S (\epsilon_{S1}, t_1), \delta_S
(\epsilon_{S2}, t_2)]g,
\end{equation}
as determined by eq. (\ref{eq:ssmalgebra}) agrees with
\begin{equation}
[\delta'_{S1}, \delta'_{S2}]g = [\delta (\epsilon_{L1}, t_1) + \tilde{\delta}
(\epsilon_{R1}, t_1), \delta(\epsilon_{L2}, t_2) + \tilde{\delta}
(\epsilon_{R2}, t_2)]g,
\end{equation}
as determined by the formulas of Section 2.  The first term in eq.
(\ref{eq:ssmalgebra}) and the $[\delta, \delta]$ and $[\tilde{\delta},
\tilde{\delta}]$ pieces
of $[\delta'_{S1}, \delta'_{S2}]$ certainly do match.
Therefore, we need to compare the second
term in eq. (\ref{eq:ssmalgebra})
with the cross terms in $[\delta'_{S1}, \delta'_{S2}]$.

According to eq. (\ref{eq:tildecom}), so painstakingly derived in Section 2.6,
the cross terms are
\[([\delta(\epsilon_{L1}, t_1), \tilde{\delta} (\epsilon_{R2}, t_2)]
+[\tilde{\delta} (\epsilon_{R1}, t_1), \delta (\epsilon_{L2}, t_2)])g \]
\begin{equation}
= {t_1 t_2\over 1- t_1 t_2} \Big(\delta (\epsilon'_{12}, t_1) + \tilde{\delta}
(g_0\epsilon'_{12} g_0^{-1}, t_2)
- \delta (\epsilon'_{21}, t_2) - \tilde{\delta} (g_0
\epsilon'_{21} g_0^{-1}, t_1)\Big)g , \label{eq:forma}
\end{equation}
where
\begin{equation}
\epsilon'_{12} = \epsilon_{L1} g_0^{-1} \epsilon_{R2} g_0 - g_0^{-1}
\epsilon_{R2} g_0 \epsilon_{L1}, \quad
\epsilon'_{21} = \epsilon_{L2} g_0^{-1} \epsilon_{R1} g_0 - g_0^{-1}
\epsilon_{R1} g_0 \epsilon_{L2}.
\end{equation}
On the other hand, the contribution to $[\delta_{S1}, \delta_{S2}]g$
from the second term in eq. (\ref{eq:ssmalgebra}) is
\begin{equation}
 - {t_1 t_2\over 1-t_1 t_2} \Big(\delta_S (\epsilon'_{S12}, t_2) - \delta_S
(\epsilon'_{S21}, t_1)\Big) M.
\end{equation}
The upper block of this gives (using eq. (\ref{eq:deltas}))
\begin{equation}
- {t_1 t_2\over 1-t_1 t_2} \Big(\delta(\epsilon'_{L12}, t_2) + \tilde{\delta}
(\epsilon'_{R12}, t_2)
-\delta (\epsilon'_{L21}, t_1) - \tilde{\delta}
(\epsilon'_{R21}, t_1)\Big) g.  \label{eq:formb}
\end{equation}
Now we must compare the expressions (\ref{eq:forma}) and (\ref{eq:formb}).
What is required for them to agree is
\begin{equation}
\epsilon'_{L12} = \epsilon'_{21} , \quad \epsilon'_{R12} = - g_0 \epsilon'_{12}
g_0^{-1},
\end{equation}
and two more relations obtained by interchanging the indices 1 and 2.  To
understand how these arise, one must substitute $M_0 = \left(\begin{array}{cc}
g_0 & 0\\ 0 & g_0^{-1}\end{array}\right)$ and $\epsilon =
\left(\begin{array}{cc} \epsilon_L & 0 \\ 0 & \epsilon_R\end{array}\right)$
into
eq. (\ref{eq:parameter}).  The upper block then gives
\begin{equation}
\epsilon'_{L12} = g_0^{-1} \tilde{\epsilon}_{L1} g_0 \epsilon_{L2} -
\epsilon_{L2} g_0^{-1} \tilde{\epsilon}_{L1} g_0.
\end{equation}
Since $\tilde{\epsilon} = - L \epsilon L$
implies that $\tilde{\epsilon}_{L1} = -
\epsilon_{R1}$, these substitutions give the desired relations, which
completes the proof.
The conclusion of this exercise is that the new term in the SSM symmetry
algebra is exactly what is required to ensure that when the PCM is formulated
as an SSM, its symmetry algebra is the usual one.

\subsection{Interpretation of the algebra}

In the preceding subsection we learned that an SSM based on $G\times G$ is
equivalent to a PCM for the group $G$.  This suggests that if the symmetry of a
PCM is the current algebra derived in Section 2,
the symmetry of an SSM should be half
as large.  The purpose of this subsection is to make this statement precise.

Let us begin by defining
\begin{equation}
\Delta_n (\epsilon) g = \int_{\cal C} {dt\over 2\pi i} t^{-n -1} \delta
(\epsilon,t) g,
\end{equation}
just as we did for PCM's.  Then it is straightforward to transcribe the algebra
in eq. (\ref{eq:ssmalgebra})
in terms of these symmetry transformations.  One finds that
(up to local $H$ transformations)
\begin{equation}
[\Delta_{m} (\epsilon_1), \Delta_{n} (\epsilon_2)] =
\Delta_{m + n} (\epsilon_{12}) + \Delta_{n - m}
(\epsilon'_{12}). \label{eq:ssmcomm}
\end{equation}
At first sight this appears to be
ambiguous. Depending on how one chooses contours, the
second term is either $\Delta_{n - m} (\epsilon'_{12})$ or
$-\Delta_{m - n} (\epsilon'_{21})$.  We will
discover, however, that they are the same, so there is no ambiguity.

It is now convenient (but not essential) to choose the gauge $M_0 = 1$.  The
significance of doing this was explained for PCM's in Section 2.4.  The same
sort of reasoning applies here.  With this choice $\epsilon'_{12} =
[\tilde{\epsilon}_1, \epsilon_2]$, which is the same as $[\epsilon_1,
\epsilon_2]$ if $\epsilon_1 \in \K$ and as $-[\epsilon_1, \epsilon_2]$ if
$\epsilon_1 \in \H$.

The mixing of the positive and negative modes implies that we
do not have a standard affine current algebra
$\hat G$, but rather a
generalization, which we propose to
call $\hat G_H$.  The idea is that the currents
\begin{equation}
J^i (\sigma) = \sum_{-\infty}^\infty J_n^i e^{in\sigma},
\end{equation}
defined on a circle, should be replaced by a pair of currents
on the semicircle (or line segment) $0 \leq \sigma \leq \pi$
\begin{equation}
J^i_N(\sigma) = J_0^i + \sum_{n=1}^\infty \cos n \sigma  (J_n^i + J_{-n}^i)
\quad {\rm and} \quad J^i_D (\sigma) =
i\sum_{n=1}^\infty \sin n \sigma (J_n^i -J_{-n}^i).
\end{equation}
The way to account for eq.(\ref{eq:ssmcomm}) is to now
impose the boundary conditions
\begin{equation}
J^{i\prime} (0) = J^{i\prime} (\pi) =0 \quad{\rm for}
\quad J^i \in {\cal H}
\quad {\rm and} \quad
J^i (0) =J^i (\pi) = 0 \quad{\rm for}\quad J^i \in {\cal K}.
\end{equation}
This means that the $\H$ currents
satisfy Neumann boundary conditions ($J_n^i = J_{-n}^i$)
and the $\K$ currents satisfy Dirichlet
boundary conditions ($J_n^i = - J_{-n}^i$)
at the two ends.  Thus the mode expansions become
\begin{equation}
J^i (\sigma) = J_0^i + 2 \sum_{n=1}^\infty \cos n \sigma
J_n^i \quad{\rm for}\quad J^i \in {\cal H}
\end{equation}
\begin{equation}
J^i (\sigma) = 2 i\sum_{n=1}^\infty \sin n \sigma J_n^i \quad{\rm for}
\quad J^i \in {\cal K}.
\end{equation}
In terms of modes, local current algebra on the line segment then implies
\begin{equation}
[J_m^i, J_n^j] = f^{ij}{}_k (J_{m + n}^k + J_{m-n}^k) \quad{\rm for}\quad
J_n^j \in {\cal H}
\end{equation}
\begin{equation}
[J_m^i, J_n^j] = f^{ij}{}_k (J_{m+n}^k - J_{m-n}^k) \quad{\rm for}
\quad J_n^j \in {\cal K},
\end{equation}
which corresponds to what we have found in eq.(\ref{eq:ssmcomm}).
In particular, we have the desired relation
$\Delta_{n - m} (\epsilon'_{12}) = -\Delta_{m - n} (\epsilon'_{21})$, which
ensures independence from how the contours are arranged in the commutator.

Note that the algebra presented here only
contains zero modes for the subalgebra $\H$, whereas we know that an SSM has a
full global $G$ symmetry.  The explanation of this is that the $\K$ zero modes
have been used up by the gauge choice $M_0 = 1$.
It is instructive to study the algebra of zero modes in some detail when $M_0$
is unrestricted to see how the separation of the $\H$ generators and the $\K$
generators is achieved.  The analysis is analogous to that given for PCM's at
the end of Section 2.4.  Let us define $\delta (\epsilon) g = g\epsilon$ and
$\bar\delta (\epsilon) = g M_0^{-1} \epsilon^{\dagger}  M_0$, and rewrite these
formulas in the more symmetrical way
\begin{equation}
\delta g = g g_0^{-1} \tilde{\epsilon} g_0 \quad
{\rm and} \quad \bar\delta g = g
g_0^{-1} \tilde{\epsilon}^{\dagger} g_0,
\end{equation}
where $\tilde{\epsilon} = g_0 \epsilon g_0^{-1}$ and $M_0 = g_0^{\dagger} g_0$.
 (This use of a tilde is different from the one in Section 5.3. We are only
considering non-compact SSM's here.) Next
we note that $\delta g_0 = \tilde{\epsilon} g_0$ and $\bar\delta g_0 =
\tilde{\epsilon}^{\dagger} g_0$, and use these equations to compute the various
commutators
\begin{equation}
[\delta(\epsilon_1), \delta (\epsilon_2)] g = g g_0^{-1} [\tilde{\epsilon}_1,
\tilde{\epsilon}_2]g_0
\end{equation}
\begin{equation}
[\delta (\epsilon_1), \bar\delta (\epsilon_2)] g = g g_0^{-1}
[\tilde{\epsilon}_2^{\dagger}, \tilde{\epsilon}_1^{\dagger}]g_0
\end{equation}
\begin{equation}
[\bar\delta (\epsilon_1), \bar\delta (\epsilon_2)] = g g_0^{-1}
([\tilde{\epsilon}_2^{\dagger}, \tilde{\epsilon}_1^{\dagger}]
+ [\tilde{\epsilon}_2^{\dagger},\tilde{\epsilon}_1] + [\tilde{\epsilon}_2,
\tilde{\epsilon}_1^{\dagger}])g.
\end{equation}

The next step is to identify the transformation that corresponds to the zero
modes of $\hat G_H$.  The correct identification turns out to be
\begin{equation}
\Delta_0 (\epsilon) = {1\over 2} \Big(\delta (\epsilon) -
\bar\delta (\epsilon) \Big),
\end{equation}
with $\tilde{\epsilon} \in \H$.  To see this, note that
\begin{equation}
[\Delta_0 (\epsilon_1), \bar\delta (\epsilon_2)] = {1\over 2} g g_0^{-1}
([\tilde{\epsilon}_1, \tilde{\epsilon}_2^{\dagger}] +
[ \tilde{\epsilon}_1^{\dagger}, \tilde{\epsilon}_2])g_0.
\end{equation}
vanishes when $\tilde{\epsilon}_1 \in \H$ and $\tilde{\epsilon}_2 \in \K$,
since then $\tilde{\epsilon}_1$ is antihermitian and $\tilde{\epsilon}_2$ is
hermitian.  In this way we have divided the transformations into $\H$
transformations $\Delta_0$ and $\K$ transformations $\bar\delta$ that commute
with one another, which is what is required.  Also note that
\begin{equation}
[\Delta_0 (\epsilon_1), \Delta_0 (\epsilon_2) ] g  =  {1\over 4} g g_0^{-1}
([\tilde{\epsilon}_1, \tilde{\epsilon}_2] + [\tilde{\epsilon}_1^{\dagger},
\tilde{\epsilon}_2^{\dagger}]
-  [\tilde{\epsilon}_1, \tilde{\epsilon}_2^{\dagger}]
- [\tilde{\epsilon}_1^{\dagger}, \tilde{\epsilon}_2])g_0.
\end{equation}
The right-hand side of this equation simplifies
to $\Delta_0 (\epsilon_{12}) g$
when $\tilde{\epsilon}_1, \tilde{\epsilon}_2 \in \H$,
as it should.

The story is completed by noting that
$[\Delta_m (\epsilon_1), \bar\delta
(\epsilon_2)] = 0$ for all $m$ when $\epsilon_1 \in \H$
and $\epsilon_2 \in \K$.
The commutator $[\bar\delta (\epsilon_1),
\bar\delta (\epsilon_2)]$ for
$\tilde{\epsilon}_1, \tilde{\epsilon}_2 \in \K$ is not readily identified.
However, whatever it is, the Jacobi identity
ensures that it too commutes with
$\Delta_m (\epsilon)$ for $\epsilon \in \H$.
The $\bar\delta$ transformations for $\K$ are just sufficient to make the gauge
choice $M_0 = 1$.  Once this is done $\Delta_0 (\epsilon)$ simplifies to
$\Delta_0 (\epsilon) g = {1\over 2} g
(\epsilon - \epsilon^{\dagger}) = g \epsilon$ for
$\epsilon \in \H$.

The relationship between the current algebra symmetries of a PCM
with group $G$ and its description as a $G\times G /G$ SSM
has a simple explanation. In the latter description, both the subgroup and the
coset currents belong to $\G$. The subgroup currents are expanded
on the interval $0\leq \sigma \leq \pi$ in cosines to implement Neumann
boundary conditions, and the coset currents are expanded in sines to
implement Dirichlet boundary conditions. Viewed as a PCM,
these are equivalent to an arbitrary Fourier series expansion
of $\G$ currents on the circle $0\leq \sigma \leq 2\pi$, as was pointed out
at the end of Section 3.3.
The discussion of zero modes presented above, when applied
to the SSM $G \times G/G$ described in Section 5.4, is equivalent to the one
presented at the end of Section 2.4 for PCM's.

\subsection{Virasoro symmetry}

It is interesting to analyze the Virasoro symmetry of SSM's.
The natural guess for the symmetry
transformation is that it is the same formula as for PCM's, namely
\begin{equation}
\delta^V (t) g = g \eta_V = g \Big((t^2 - 1) \dot X (t) X (t)^{-1} + \I\Big).
\label{eq:virsym}
\end{equation}
Using eqs. (\ref{eq:aformula}) and (\ref{eq:deltaamu})
\begin{equation}
\delta^V A_\mu = D_\mu \eta_V + M^{-1} \partial_\mu \eta_V^\dagger M.
\end{equation}
We already know from
the study of PCM's that the first term preserves $\partial \cdot A = 0$, so the
only thing left to check is whether the second one does too.

To study $\partial^\mu (M^{-1} \partial_\mu \eta_V^\dagger M)$, we first derive
\begin{equation}
M^{-1} \partial_{\pm} \eta_V^\dagger M
= ((1 - t^2) \dot\alpha_\pm \pm 1) A_\pm +
\alpha_\pm (t^2-1) [ A_\pm, M^{-1} (\dot X X^{-1})^\dagger M].
\end{equation}
Then, using $\partial_- A_+ = - \partial_+ A_- = {1\over 2} [ A_+, A_-]$, we
obtain
\[\partial^\mu (M^{-1} \partial_\mu \eta_V^\dagger M) = \Big(1 + (t^2-1)
(-{1\over 2} \dot\alpha_+ + {1\over 2} \dot\alpha_- + \alpha_+
\dot\alpha_- + \alpha_- \dot\alpha_+)\Big)[A_+, A_-]\]
\begin{equation}
 + (t^2 - 1) (\alpha_+ \alpha_- - {1\over 2} \alpha_+ - {1\over 2} \alpha_-)
\Big[ [A_+, A_-], M^{-1} (\dot X X^{-1})^\dagger M\Big].
\end{equation}
The coefficients of both terms are zero, so the symmetry is verified.

The conserved charges associated with the hidden Virasoro symmetry of SSM's
contain the result given for PCM's in eq. (\ref{eq:pcmvircharge}) plus an
additional contribution
given by
\begin{equation}
\int_{-\infty}^\infty D_0 (M^{-1} ((t^2 - 1) (\dot X X^{-1})^\dagger + \I) M
dx^1 = {1\over t} \int_{-\infty}^\infty \partial_1((t^2 - 1) M^{-1}
(\dot X X^{-1})^\dagger M - \I) dx^1.
\end{equation}
Adding this to the expression in eq. (\ref{eq:pcmvircharge}) gives the result
\begin{equation}
Q^V (t) = \Big(\J + {t^2-1\over t} (\dot X X^{-1} + M^{-1} (\dot X
X^{-1})^\dagger M)\Big) \Big|_{-\infty}^\infty.
\end{equation}

The more challenging calculation is the algebra of commuting a Virasoro
symmetry with a current algebra symmetry.  In others words, we wish to evaluate
\begin{equation}
[\delta_1^V, \delta_2] g = [\delta^V (t_1), \delta (\epsilon_1, t_2)] g.
\end{equation}
using eqs. (\ref{eq:virsym}) and (\ref{eq:casym}).
In the first instance, the commutator is
\[ [\delta_1^V, \delta_2] g  =  g [(t_1^2 - 1) \dot X_1 X_1^{-1} + \I, X_2
\epsilon X_2^{-1}]  +  g \delta_1^V (X_2 \epsilon X_2^{-1}) \]
\begin{equation}
 - g \delta_2 \Big((t_1^2 - 1) \dot X_1
X_1 + \I\Big).
\end{equation}
To make further progress, we need convenient expressions for $\delta_1^V X_2$,
$\delta_2 X_1$, and $\delta_2 \I$.  The formula for $\delta_2 X_1$ has already
been obtained in eq. (\ref{eq:dxssm}).
$\delta_1^V X_2$ was found in eqs. (\ref{eq:dxvira}) and (\ref{eq:dxvirb})
for PCM's. However, now it is expected to acquire additional terms.
The appropriate formulas, obtained by the usual methods, are
\begin{equation}
\delta_2 X_1 = {t_1\over t_2 - t_1} (\eta_2 - \eta_1) X_1 + \lambda (M^{-1}
\eta_2^\dagger M X_1 - X_1 M_0^{-1} \epsilon^\dagger M_0)
\end{equation}
\begin{equation}
\delta_2 I = {1\over t_2} (\eta_2 - \epsilon) + t_2 (M^{-1} \eta_2^\dagger
M - M_0^{-1} \epsilon^\dagger M_0)
\end{equation}
\[ \delta_1^V X_2 =(\gamma_a \dot X_1 X_1^{-1} + \gamma_c \dot X_2 X_2^{-1} -
I)X_2  - \lambda (1 - t_2^2) \dot X_2 \]
\begin{equation}
- \lambda (1 - t_1^2) M^{-1} (\dot X_1 X_1^{-1})^\dagger M X_2,
\end{equation}
where $\eta_i = X_i \epsilon X_i^{-1}, \gamma_a$ and $\gamma_c$ are given in
eq. (\ref{eq:dxvirb}), and $\lambda$ is given in eq. (\ref{eq:lambdaeq}).

Using the formulas above we find that
\begin{equation}
[\delta_1^V, \delta_2] g = \delta g + \delta' g + \delta^{\prime\prime} g,
\end{equation}
where $\delta g$ is the result obtained for PCM's
in eq. (\ref{eq:deltagpcm}), $\delta' g$ is
a local $H$ transformation which may be dropped, and $\delta^{\prime\prime} g$
contains new terms which must be kept.  Explicitly,
\[ \delta g =  \Big({1\over t_2}
(\delta (\epsilon, 0) - \delta
(\epsilon, t_2)) + {t_2 (t_1^2 - 1)\over (t_1 - t_2)^2} (\delta (\epsilon, t_1)
- \delta (\epsilon, t_2))\]
\begin{equation}
+ {t_1 (1 - t_2^2)\over t_1 - t_2} ~ {\partial\over\partial t_2} \delta
(\epsilon, t_2) \Big) g.
\end{equation}
\[ \delta' g  =  (1 - t_1^2) \lambda (g^\dagger)^{-1}
\Big(\eta_2^\dagger M \dot
X_1 X_1^{-1} - (X_1 X_1^{-1})^\dagger M \eta_2 \Big) g^{-1} \]
\[ +  (1 - t_1^2)\lambda g (\eta_2 M^{-1}
(\dot X_1 X_1^{-1})^\dagger - \dot X_1 X_1^{-1}
M^{-1} \eta_2^\dagger) g \]
\begin{equation}
+ (1 - t_1^2){\partial\over\partial t_1} \Big(\lambda ((g^\dagger)^{-1}
\eta_2^\dagger - g \eta_2 g ^{-1})\Big)
- {t_2} \Big((g^\dagger)^{-1} \eta_2^\dagger g^\dagger - g
\eta_2 g^{-1})\Big) g
\end{equation}
and
\[ \delta^{\prime\prime} g = g \Big( (1 - t_1^2)
{\partial \lambda\over \partial
t_1} \eta_2 - t_2 \eta_2 - (1 - t_2^2) \lambda {\partial\eta_2\over\partial
t_2}   + t_2 M_0^{-1} \epsilon^\dagger M_0\]
\begin{equation}
 - (1 - t_1^2) {\partial\lambda\over\partial t_1} X_1 M_0^{-1} \epsilon^\dagger
M_0X_1^{-1}\Big).
\end{equation}
The crucial question becomes what $\delta^{\prime\prime} g$ contributes to
$[\delta_m^V, \delta_n (\epsilon)]g$, when we insert it into the appropriate
contour integrals, or, equivalently, what it contributes to $[K_m, J_n^i]$.
Again, depending on how the contours are arranged, the full answer is given by
the first three terms or by the last two terms in $\delta^{\prime\prime}$.
Either way, one obtains the same result, namely
\begin{equation}
[K_m, J_n^i] = n (J_{m+n-1}^i - J_{n+n-1}^i - J_{n-m+1}^i + J_{n - m+1}^i).
\end{equation}
The first two terms are the result obtained previously
in eq.(\ref{eq:pcmalg}) from $\delta g$, while
the last two terms are the new contribution arising from $\delta^{\prime\prime}
g$.

The interpretation of this result is evident.  The right-hand side is invariant
under $m \rightarrow - m$.  Therefore, the generators $K_m$ satisfy the
restrictions $K_m = K_{-m}$, just like the $\H$ currents.  In other words,
$K(\sigma)$ satisfies Neumann boundary conditions at $\sigma = 0$ and $\sigma =
\pi$.  Just as in Section 3.3, the internal stress tensor
\begin{equation}
T(\sigma) = {i\over 2} {K(\sigma)\over \sin \sigma}
\end{equation}
satisfies the standard stress tensor algebra, but it is singular at $\sigma =
0$ and $\sigma = \pi$.  It is interesting that the boundary condition takes a
simpler form for $K(\sigma)$ than it does for $T(\sigma)$.

In Section 5.5 we learned how to relate the current algebra symmetries of the
$G \times G/G$ SSM to those of the equivalent PCM.  Now let's consider the
same question for the Virasoro symmetries.  In the PCM formulation the
currents $K(\sigma)$ on the circle can be replaced by a pair of currents
on the semicircle, one
satisfying Neumann boundary conditions and one satisfying Dirichlet boundary
conditions.  These are the cosine and sine terms in the Fourier series
expansion.  On the other hand, for a general SSM we obtained a symmetry current
$K(\sigma)$ satisfying Neumann boundary conditions.  This is all that should
have been expected, since when commuted with the $\H$ and $\K$ currents this is
what is required to preserve their boundary conditions.  However, this raises
the question what happened to the Dirichlet part of the PCM Virasoro symmetry.
There is only one possible answer that makes sense.  The $G \times G/G$ SSM
must have an additional symmetry current $\tilde{K}(\sigma)$, satisfying
Dirichlet boundary conditions, which when commuted with $\H$ currents gives
$\K$ currents and vice-versa.  This is possible in this particular case,
because both the $\H$ and $\K$ currents correspond to $G$.  Clearly, this is a
special feature of the $G\times G/G$ model, and not a general feature of SSM's.

\sectiono{Conclusion}

\subsection{Summary of main results}

This paper has presented a study of hidden symmetries for certain classical
theories in two-dimensional Minkowski space.  Two classes of theories were
considered -- principal chiral models (PCM's), for which the basic variables
$g(x)$ provide a map of the space-time into the group manifold of a compact
Lie
group $G$, and symmetric space models (SSM's), for which $g(x)$ maps space-time
into a symmetric space $G/H$, where $G$ is a non-compact Lie group and $H$ is
its maximal compact subgroup.

In the case of PCM's it has been known for a long time that
the hidden symmetry is
$\hat G$,  the Kac--Moody (or loop group) extension of $G$.  Our analysis
confirmed this result.  In the case of SSM's, it has been generally believed
(with one notable exception \cite{bib:wu})
that the hidden symmetry is also $\hat G$.
Section 5 showed that this is not the case, but rather that the hidden symmetry
is given by a subalgebra of $\hat G$, which we called $\hat G_H$.  The
interpretation of $\hat G_H$ is that instead of having a current algebra on a
circle $ 0\leq \sigma < 2 \pi$, there is one on an interval $0 \leq \sigma \leq
\pi$.  Furthermore, the currents belonging to the subalgebra $\H$ satisfy
Neumann boundary conditions at the ends of the interval (so that the modes
satisfy $J_{-m}^i = J_m^i$) and those of the coset $\K$ satisfy Dirichlet
boundary conditions (so that $J_{-m}^i = - J_m^i$).  These results were shown
to
be compatible with the formulation of a PCM for the group $G$ as an SSM for the
coset $(G \times G)/ G$.

Virasoro symmetries were also studied.  The symmetry group was found to
correspond to a subalgebra of the Virasoro algebra, which we called $\V_2$.
It was proved that this could not be extended to the full Virasoro algebra
$\V$, because the hidden symmetry stress tensor $T (\sigma$) is singular at the
ends of the interval.  In the case of SSM's it turned out that the $\V_2$
current $K(\sigma)$ satisfies Neumann boundary conditions at the ends of the
interval.

The structure of the hidden symmetries suggested exploring a particular change
of variables, which was interpreted as implementing a non-abelian duality
transformation depending on a continuous parameter $u$.  The $u$-transformed
theory turned out to correspond to a PCM with a Wess--Zumino term added with
coefficient $u$.  This implied at least classically, that the theory with a WZ
term is the same as the theory without one, provided only that $u\not= \pm 1$,
which would give WZNW theory.
It was hinted that interesting additional issues arise at the
quantum level.  The hidden symmetry of the classical $u$-transformed theory was
constructed directly as a straightforward generalization of the construction in
the $u = 0$ case.

\subsection{Future directions}

There are a number of directions in which the present work should be extended,
some of which are expected to be relatively straightforward to analyze.  For
the string theory program that was sketched in the introduction, the direction
to go is clear.  First of all, the models studied here should be coupled to
gravity.  This is currently being investigated \cite{bib:schwarza}.
Next we should specialize to the specific models of greatest relevance to
string theory.  These are the SSM's based on $O(8, 24)$ and $E_{8,8}$, which,
when the fermions are added, have $N = 8$ and $N = 16$ supersymmetry,
respectively.  It will be interesting to learn whether the inclusion of
fermions
and supersymmetry leads to additional hidden symmetries as suggested by
Nicolai \cite{bib:nicolaia}. The quantum effects that break the symmetry groups
to discrete subgroups should also be studied.

Another possible direction, which
could prove to be quite non-trivial, is to repeat the
analysis of this paper with a spatial dimension that is a circle rather than a
line.  Once this has been successfully carried out, it will be possible to
truncate to the zero modes to define the most relevant and promising reduction
to one dimension.\footnote{ A different reduction to one
dimension was considered in ref. \cite{bib:nicolaic}.}
It is here that one hopes to find symmetries given by
hyperbolic algebras.  If this is correct, it will be fascinating to understand
in detail how they are realized.  Maybe then will we have a better
understanding of the hidden gauge symmetries of string theory.

There are other directions worthwhile exploring that are relevant to issues
different from the goal of identifying the gauge symmetries of fundamental
strings.  For example, there is much that can still be done towards better
understanding the quantum behavior of these two-dimensional models.  Polyakov's
original vision of using them as toy models of four-dimensional gauge theories
could still bear new fruit.  It would also be interesting to bring other
integrable models into the same general framework.

\subsection{Acknowledgements}

It is a pleasure to express my gratitude to Ashoke Sen, with whom I have had
frequent communications throughout the course of this work.  Also, he has kept
me apprised of the status of his closely related research and offered probing
questions and comments on my work.  I also wish to acknowledge stimulating
discussions with L. Chau, S. Cherkis, A. Dabholkar, J. Gauntlett,
G. Gibbons, C. Johnson, J. Maharana, M. Perry, and A. Schwarz.
Thanks also go to D. Maison and S. Rajeev for
sending me hard-to-find preprints \cite{bib:maisona,bib:rajeev}
and to M. Green and C. Hull for reading the
manuscript and offering helpful suggestions.

\end{document}